\documentclass{amsart}
\usepackage{cite}
\usepackage{pst-sigsys}
\usepackage[utf8]{inputenc}
\usepackage{amsmath}
\usepackage{amssymb}
\usepackage{amscd}
\usepackage{amsfonts}
\usepackage{amsthm,color,graphicx}

\usepackage{booktabs}
\usepackage[caption=false,font=normalsize,labelfont=sf,textfont=sf]{subfig}
\usepackage{url}

\newcommand{\bd}{\mathbf}
\newcommand{\bdi}{\boldsymbol}
\def\lcm{\operatorname{lcm}}
\def\RR{\mathbb{R}}
    
    \def\ZZ{\mathbb{Z}}
    
    \def\NN{\mathbb{N}}
    
\theoremstyle{plain}
\newtheorem{Thm}{Theorem}

\theoremstyle{definition}

\begin{document}

\title[A Perceptually Motivated Filter Bank with Perfect Reconstruction]{A Perceptually Motivated Filter Bank with Perfect Reconstruction for Audio Signal Processing}

\author{Thibaud~Necciari, Nicki~Holighaus, Peter~Balazs and Zden\v{e}k~Pr\r{u}\v{s}a}
 \address{Acoustics Research Institute Austrian Academy of Sciences, Wohllebengasse 12--14, A-1040 Vienna, Austria}
 \email{\{thibaud.necciari,nicki.holighaus,peter.balazs,zdenek.prusa\}@oeaw.ac.at}

\begin{abstract}

Many audio applications rely on filter banks (FBs) to analyze, process, and re-synthesize sounds. To approximate the auditory frequency resolution in the signal chain, some applications rely on perceptually motivated FBs, the gammatone FB being a popular example. However, most perceptually motivated FBs only allow partial signal reconstruction at high redundancies and/or do not have good resistance to sub-channel processing. This paper introduces an oversampled perceptually motivated FB enabling perfect reconstruction, efficient FB design, and adaptable redundancy. The filters are directly constructed in the frequency domain and linearly distributed on a perceptual frequency scale (e.g. ERB, Bark, or Mel scale). The proposed design allows for various filter shapes, uniform or non-uniform FB setting, and large down-sampling factors. For redundancies $\geq$ 3 perfect reconstruction is achieved by computing the canonical dual FB analytically. For lower redundancies perfect reconstruction is achieved using an 
iterative method. Experiments show performance improvements of the proposed approach when compared to the gammatone FB in terms of reconstruction error and resistance to sub-channel processing, especially at low redundancies.

\end{abstract}

\maketitle

\section{Introduction}

Time-frequency (TF) transforms like the short-time Fourier or wavelets transforms play a major role in audio signal processing. They allow decomposing any signal into a set of elementary functions with good TF localization and achieving perfect reconstruction if the transform parameters are chosen appropriately (e.g. \cite{Flandrin:1999a}). Therefore, they constitute ideal tools to analyze, process and re-synthesize sounds. Accordingly, applications like audio coding \cite{Baumgarte:2002a,Strahl:2009a}, audio transformations \cite{sirdey:hal-00881707,ltfatnote022}, sparsity \cite{Balazs:2010a,papkow13}, source separation \cite{6334422,Gao:2014a}, speech processing \cite{Gunawan:2010a,maba09}, de-noising \cite{Donoho:1995a,majxxl10}, or optimization of acoustical measurements \cite{majxxl1}, among others, rely on TF decompositions to perform sub-channel processing and reconstruct the signal from the modified TF components. In such applications, TF transforms are usually implemented as filter 
banks (FBs) where the set of analysis filters defines the elementary functions and the set of synthesis filters allows for signal reconstruction. The TF concentration of the filters together with the downsampling factors in the sub-bands define the TF resolution and redundancy of the transform. FBs come in various flavors and have been extensively treated in the literature (e.g. \cite{Akkarakaran:2003a,Vaidyanathan:1993a}). Note that the mathematical theory of frames constitutes an interesting alternative background for the interpretation and implementation of FBs (see e.g.~\cite{Bolcskei:1998a,Cvetkovic:1998a,Balazs:2011a,Fickus:2013a,badokowto13} and Appendix~\ref{sec:nonuniformFB}).

Because sub-channel processing may introduce audible distortions in the reconstructed signal, particularly if the sub-bands are not equally processed, important requirements of audio applications include: a {\em strong} stability (i.e. the coefficients are bounded if and only if the signal is bounded, i.e. the FB and its inverse are BIBO-stable), perfect reconstruction property of the analysis-synthesis system (i.e. when no sub-channel processing is performed)\footnote{Note that because of the quantization process, the perfect reconstruction requirement might be violated for lossy coding applications.}, resistance to noise, and adequate aliasing suppression in each sub-band. To limit the computational costs, certain applications also require low redundancy, that is a small number of sub-bands with large downsampling factors. While sub-sampling the sub-bands is usually not required in speech processing where signals are often short and sampled at low rates (typically 8~kHz), sub-sampling is of high interest 
to music processing where signals are few seconds long and sampled at high rates ($\geq$ 44.1~kHz). 

Although many applications still use transforms with fixed resolution (e.g. short-time Fourier or modified cosine transforms), there is a strong desire in audio processing to analyze sounds in a manner similar to that of the human ear. Since the auditory TF resolution varies with frequency, this implies using a transform with variable resolution. This purpose has lead to the design of so-called auditory FBs (e.g. \cite{Lopez-Poveda:2001a,Hohmann:2002a,Irino:2006b}) or perceptually motivated FBs (e.g. \cite{Cvetkovic:2003a,Schoerkhuber:2014a,Venkitaraman:2014a}). Perceptually motivated FBs are usually intended as signal processing tools and, as such, they are linear, partially invertible, and have good aliasing suppression but only approximate the auditory frequency resolution. In contrast, auditory FBs are usually intended as perceptual analysis tools and, as such, they attempt to reproduce the nonlinear processing in the auditory system, to the detriment of perfect reconstruction, aliasing suppression, and 
computational efficiency. A \textit{linear} and partially invertible auditory FB became popular in audio signal processing, though, namely the gammatone FB (e.g. \cite{Patterson:1992a,Hohmann:2002a}). Gammatone filters approximate well the auditory TF resolution at low to moderate sound levels and are easy to implement as FIR or IIR filters \cite{Lin:2001b,Hohmann:2002a,Lyon:2010a}. A wide range of audio applications thus implements gammatone FBs, for instance source separation \cite{Gao:2014a}, speech processing \cite{Gunawan:2010a,Zhao:2012a,Sadjadi:2015a}, or music information retrieval \cite{Valero:2012a}. Still, gammatone FBs do not satisfy all requirements of audio applications as they neither provide perfect reconstruction nor good aliasing suppression (see Sec.~\ref{ssec:auditoryFB}).

To fulfill the requirements of audio applications, this paper introduces an oversampled perceptually motivated FB enabling perfect reconstruction, efficient FB design, and adaptable resolution and redundancy. In the proposed approach, the filters are directly constructed in the frequency domain and linearly distributed on a perceptual frequency scale (e.g.~ERB, Bark, or Mel scale). The proposed design allows for various filter shapes, uniform or non-uniform FB setting, and large downsampling factors. For redundancies $\geq$~3 perfect reconstruction is achieved by computing the dual FB directly. For lower redundancies (down to $\approx$1.1) perfect reconstruction is achieved using an iterative method. Experiments show the better performance of the proposed approach with respect to the commonly-used gammatone FB in terms of reconstruction error and resistance to sub-channel processing, especially at low redundancies. 

The paper is organized as follows. The next section briefly describes the properties of auditory frequency selectivity and perceptual frequency scales, and reviews recent works related to the present study. Section~\ref{sec:erbfb} introduces the analytical and implementation properties of the proposed ``AUDlet'' FB. Finally, simulations are performed in Section~\ref{sec:simu} to show the performance of the AUDlet FB in terms of signal representation, reconstruction error, and as an audio processing tool. In the appendix, general results on non-uniform FBs and their connection to the theory of frames are recalled for the better understanding of the AUDlet FB construction.

\section{Background}

\subsection{\label{ssec:audres}Auditory Frequency Selectivity}

The frequency selectivity of the auditory system can be modeled in a first approximation as a bank of bandpass filters, named ``critical bands'' or ``auditory filters'', that are related to the frequency-to-place transformation in the cochlea (see e.g.~\cite[Chap.~3]{Moore:2012a} for a review). Briefly, when a sound reaches the ear it produces a vibration pattern on the basilar membrane. The position and width of this pattern along the membrane depend on the spectral content of the sound. Accordingly, the center frequency and bandwidth of the auditory filters respectively approximate the place and width of excitation on the basilar membrane. Noteworthy, the width of excitation depends on level as well: patterns become wider and asymmetric as sound level increases (e.g.~\cite{Glasberg:1990a}). Several auditory filter models have been proposed based on the results from masking experiments \cite{Lyon:2010a}. A popular auditory filter model is the gammatone filter \cite{Patterson:1992a}. Although gammatone 
filters do not capture the level dependency of the actual auditory filters, their ease of implementation in the time or Laplace domains made them popular in audio signal processing (e.g.~\cite{Gao:2014a,Gunawan:2010a,Valero:2012a,Zhao:2012a}). More realistic auditory filter models are, for instance, the roex and gammachirp filters \cite{Glasberg:1990a,Unoki:2006a}.

\subsection{\label{ssec:audscales}Perceptual Frequency Scales}

To analyze sounds using a frequency resolution that mimics that of the ear or to match the spectral content of a sound to an auditory sensation (e.g. pitch or loudness), a mapping between the linear frequency domain and the nonlinear perceptual domain is required. This mapping is provided by perceptual frequency scales developed based on psychoacoustics experiments. We mention below three scales that are commonly used in hearing science and audio signal processing, namely the Bark, ERB, and Mel scales. To describe the different mappings we introduce the function $F: \xi \rightarrow \mathrm{AUD}$ where $\xi$ is frequency in Hz and $\mathrm{AUD}$ is an auditory unit that depends on the scale.

\subsubsection{The Bark Scale}

Directly originates from the critical bands' concept. An expression for the Bark rate is~\cite{Zwicker:1980a}
\begin{equation}\label{eq:Barkrate}
	\begin{split}
		\mathrm{AUD}_{\mathrm{Bark}} &= F_{\mathrm{Bark}}(\xi)\\
		 & =  13 \arctan (0.00076 \xi) + 3.5 \arctan (\xi / 7500)^2 \:.
	\end{split}
\end{equation}
\noindent$\mathrm{AUD}_{\mathrm{Bark}}$  corresponds to the critical band rate expressed in Barks. The corresponding bandwidth in Hz is
\begin{equation}\label{eq:Barkbw}
	BW_{\mathrm{Bark}} = 25 + 75 \left( 1 + 1.4\times10^{-6} \xi^2\right)^{0.69} \:.	
\end{equation}

\subsubsection{The ERB Scale}

Follows the same concept as the Bark scale but results from a different set of experiments (see e.g.~\cite{Moore:2012a} for a comparison of the two scales and their underlying assumptions). The ERB rate is~\cite{Glasberg:1990a}
\begin{equation}\label{eq:erbrate}
	\mathrm{AUD}_{\mathrm{ERB}} = F_{\mathrm{ERB}}(\xi) = 9.265 \ln \left( 1 + \frac{\xi}{228.8455}\right)
\end{equation}
and the corresponding bandwidth in Hz is
\begin{equation}\label{eq:erbw}
	BW_{\mathrm{ERB}} = 24.7 + \frac{\xi}{9.265} \quad .
\end{equation}

\textit{Remark:} $BW_{\mathrm{Bark}}$ and $BW_{\mathrm{ERB}}$ are commonly used in psychoacoustics and signal processing to approximate the auditory frequency resolution at low to moderate levels (i.e. 30--70~dB) where the auditory filters' shape remains symmetric and constant. See for example \cite{Glasberg:1990a,Unoki:2006a} for the variation of $BW_{\mathrm{ERB}}$ with level.

\subsubsection{The Mel Scale}

Provides a means to quantify pitch, that is the perceived height of a note, as a function of frequency. A popular formula for the Mel scale is~\cite{Oshaughnessy:1987a}
\begin{equation}\label{eq:melrate}
	\mathrm{AUD}_{\mathrm{Mel}} = F_{\mathrm{Mel}}(\xi) = 2595 \, \log_{10} \left( 1+\frac{\xi}{700} \right) \quad .
\end{equation}
The resulting pitch value has the unit ``mel'' (as in \textit{mel}ody). Because the Mel scale is not directly related to the auditory filters' concept, it provides no expression for the bandwidth. However, it is common practice to construct Mel FBs with filters linearly distributed on the Mel scale. Their bandwidth is usually set to reach 50\% overlap between channels. This is done, for instance, to compute the so-called ``Mel-Frequency cepstrum coefficients'' (MFCCs) in the fields of speech recognition \cite{Shriberg:2007a} or music information retrieval \cite{Valero:2012a}.

\subsection{\label{ssec:auditoryFB}Related Work}

A wide variety of tools is available for achieving a perceptually motivated TF transform of a sound. Nonetheless, only a number of these tools allows to (approximately) reconstruct the signal from the transform coefficients. Many analysis-synthesis systems have been proposed that implement gammatone filters in the analysis stage and their time-reversed impulse responses in the synthesis stage (e.g.~\cite{Lin:2001b,Hohmann:2002a,Strahl:2009a,Gunawan:2010a,Zhao:2012a}). This setting implies that the frequency response of the gammatone FB has an all-pass characteristic and features no ripple (equivalently in the frame context, that the system is tight, see Appendix~\ref{sec:nonuniformFB}) while in practice it does not. A reason for that is that gammatone FBs usually consider only a limited range of frequencies (typically in the interval 0.1--4~kHz for speech processing). Therefore, such systems only achieve an approximate reconstruction, still the audio quality of the reconstruction is good provided a rather 
high density of filters is used~\cite{Lin:2001b,Strahl:2009a,Gunawan:2010a}. Moreover, commonly-used 4-th order gammatone filters (an order of 4 allows to best approximate the auditory filters' shape at low to moderate levels \cite{Patterson:1992a}) do not feature a steep decay in the frequency domain and, therefore, might not have a good aliasing suppression property (this is assessed in Sec.~\ref{sec:simu}).

Other popular tools like auditory models are motivated by the idea to replicate the nonlinear processing in the auditory system (e.g.~\cite{Lopez-Poveda:2001a,ODonovan:2005a,Irino:2006b,Zilany:2006a}). Such models are useful to improve our knowledge about the auditory system but they are generally intended for signal analysis only. Thus, they do not allow for signal reconstruction. Note that the approach proposed in \cite{Irino:2006b} does feature a re-synthesis option. This analysis-synthesis system implements compressive gammachirp filters. Nevertheless, because the synthesis stage uses the time-reversed gammachirp impulse responses, the reconstruction is only approximate and a rather large density of filters is required to achieve a good quality, as for gammatone filters.

Other tools include FB constructions aimed at mimicking the auditory frequency resolution for audio signal processing purposes. For instance, wavelet and constant-Q FBs (e.g.~\cite{Holighaus:2013a,Schoerkhuber:2014a,Venkitaraman:2014a}) are often used but they mismatch the auditory frequency resolution at low frequencies ($<$ 2~kHz, see e.g.~\eqref{eq:erbw}), where they also unnecessarily feature a large number of filters, and do not always achieve perfect reconstruction. Other approaches include FBs made of blocks of uniform frequency resolution FBs \cite{Baumgarte:2002a,Cvetkovic:2003a}. The approach in \cite{Baumgarte:2002a} only roughly approximates the ERB scale and, being targeted at audio coding, is not invertible. The approach in \cite{Cvetkovic:2003a} is designed to maximize resistance to sub-channel processing (i.e. filters with a high attenuation outside the passband) and allows for nearly perfect reconstruction for redundancies $\geq$ 2. However, since one has to properly design the windows and 
transition filters so that the desired FB properties hold, the global FB design turns out to be complex.

\section{\label{sec:erbfb} Proposed Approach: The AUDlet Filter Bank}

The present section introduces a perfect reconstruction oversampled FB that approximates the auditory frequency resolution and provides adaptable redundancy. To allow for a simple and flexible FB design, the FB construction is directly performed in the frequency domain and any compactly supported (e.g. FIR) window is an eligible filter's shape. It has to be considered, however, that the proposed concept is a \textit{linear} FB and, as such, it does not constitute an attempt to reproduce the actual \textit{non}linear auditory filtering. This is discussed below.

\subsection{Notation and Definitions}

In the following, we consider signals in $\ell_2(\ZZ)$ sampled at the frequency $\xi_s$. The inner product of two signals $x, y$ is $\left<x, y \right> =  \sum_{n} x[n] \cdot y[n]$ and the energy of a signal is defined from the inner product as $||x|| = \left<x, x\right>$. We denote the $z$-transform by $\mathcal{Z}: \, x[n] \mapsto X(z)$. By setting $z = e^{2i\pi\xi}$ for $\xi \in \mathbb{T}:=\RR/\ZZ$, the $z$-transform equals the discrete-time Fourier transform ($\mathrm{DTFT}$). Throughout the paper, bold italic letters indicate matrices (upper case), e.g. {$\bdi G$}, and vectors (lower case), e.g. {$\bdi h$}.

The proposed AUDlet FB has a general non-uniform structure as presented in Fig.~\ref{sfig:nonuniformFB} with analysis filters $H_{k}(z)$, synthesis filters $G_{k}(z)$, and downsampling and upsampling factors $d_k$. Since we consider signals in $\RR$ we deal with symmetric $\mathrm{DTFT}$s, which allows us to process only the positive-frequency range. Therefore, the letter $K$ denotes the number of filters in the frequency range $[\xi_{\mathrm{min}},\xi_{\mathrm{max}}]\cap [0,\xi_{s}/2[$, where $\xi_{\mathrm{min}} \geq 0 $ to $\xi_{\mathrm{max}} \leq \xi_{s}/2$ and $\xi_{s}/2$ is the Nyquist frequency. If $\xi_{\mathrm{min}} > 0 $, $K$ includes an additional filter at the zero frequency. Furthermore, another filter is always positioned at the Nyquist frequency. Assuming $\xi_{\mathrm{min}} = 0 $ and $\xi_{\mathrm{max}} = \xi_{s}/2$ for the rest of this manuscript, this implies that all non-uniform FBs treated below feature $K+1$ filters in total and their redundancy is defined as $R = d_{0}^{-1} + 2 \sum_{k = 
1}^{K-1} d_{k}^{-1} + d_{K}^{-1}$, since coefficients in the $1$st to $K$-th subbands are complex-valued.% An analogous formula describe the redundancy in the case $\xi_{\mathrm{min}} > 0 $.

We describe below the analysis and synthesis stages of the AUDlet FB and specify the perfect reconstruction conditions for different sets of downsampling factors $d_k$'s. To allow for FB inversion, the analysis FBs described below are constructed such that they always form a frame, i.e. with sufficiently small downsampling factors. For results regarding suitable choices of downsampling factors and references regarding the inversion (\emph{perfect reconstruction conditions}) of non-uniform FBs we refer to the appendix. By default, the algorithms referenced in this manuscript (see Sec.~\ref{ssec:impl}) automatically determine suitable downsampling factors $d_k$'s.

\subsection{Analysis Filter Bank}

The AUDlet filters $H_k$'s, $k \in \{0,\ldots,K\}$ are constructed in the frequency domain by
\begin{equation}\label{eq:erbfilters}
	H_k(e^{2i \pi\xi}) = \Gamma_k^{-\frac{1}{2}} w \left( \frac{\xi - \xi_k}{\Gamma_k} \right)
\end{equation}
where $w(\xi)$ is assumed to be a prototype filter's shape with bandwidth $1$ and center frequency $0$. This implies that the shape factor $\Gamma_k$ controls the effective bandwidth of $H_k$ and $\xi_k$ determines its center frequency. The factor $\Gamma_k^{-\frac{1}{2}}$ ensures that all filters (i.e. $\forall~\xi_k$) have the same energy. To obtain filters equidistantly spaced on a perceptual frequency scale, the sets $\{\xi_k\}$ and $\{\Gamma_k\}$ are calculated using the corresponding $\mathrm{AUD}_{\mathrm{scale}}$ and $BW_{\mathrm{scale}}$ formulas. For instance, linearly distributing $K$ filters from $\mathrm{AUD}_{\mathrm{ERB}_\mathrm{min}} = F_{\mathrm{ERB}}(\xi_\mathrm{min})$ to $\mathrm{AUD}_{\mathrm{ERB}_\mathrm{max}} = F_{\mathrm{ERB}}(\xi_\mathrm{max})$ with a density of $V$ filters per ERB leads to an ERB step $\mathrm{AUD}_{\mathrm{ERB}_k} = \mathrm{AUD}_{\mathrm{ERB}_\mathrm{min}} + k / V $. Then $\xi_k = F^{-1}_{\mathrm{ERB}}(\mathrm{AUD}_{\mathrm{ERB}_k})$ and $\Gamma_k = BW_{\mathrm{ERB}}
(\xi_k)$. Overall, the resolution of the analysis is given by two parameters: $K = V \left( \mathrm{AUD}_{\mathrm{ERB}_\mathrm{max}} - \mathrm{AUD}_{\mathrm{ERB}_\mathrm{min}} \right)$ and the set of downsampling factors $\{d_k\}$. An analogous process yields an FB adapted to any frequency scale.

\subsection{Synthesis}

In general, the existence of a non-uniform dual FB having the same number of filters $G_k$'s and upsampling factors $d_k$'s as the non-uniform analysis FB cannot be guaranteed. Therefore, we use three different approaches to compute the action of the AUDlet synthesis FB: 
\begin{enumerate}
	\item[(i)] For band-limited filters with sufficiently dense sampling, dual synthesis filters can be explicitly and efficiently computed. Synthesis is then accomplished by a standard non-uniform FB synthesis algorithm. The formal conditions for this setting are given in Thm~\ref{thm:painless} in the appendix. The dual FB is computed by \eqref{eq:candualPL} also given there.
	\item[(ii)] If the conditions of Thm~\ref{thm:painless} are violated but $\sum_{k=0}^{K} q_k$ is small enough, then the equivalent uniform FB for $H_k, d_k, \, k \in \{0,\ldots,K\},$ is constructed as described in the appendix, see Fig.~\ref{sfig:equniformFB}. A dual FB can be easily obtained using standard algorithms for the computation of dual uniform FBs.
	\item[(iii)] If the number of channels in the equivalent uniform FB is too large, the computation and storage of the dual FB become unfeasible. In such cases, the action of the canonical dual FB is computed using a conjugate gradients (CG) algorithm. Iterative synthesis via CG benefits from the fact that although the number of iterations necessary to achieve the desired precision depends on the actual frame bound ratio of the analysis FB, it does not require explicit estimates of the frame bounds as opposed to other iterative approaches like the classical frame algorithm~\cite{Grochenig:1993a}. Furthermore, since each iteration computes the analysis followed by synthesis with the filters $H_k$'s, see \eqref{eq:CG}, the algorithm's complexity is independent of the structure of the dual FB. Additionally, we showed in \cite{Necciari:2013a} that using a preconditioner often drastically reduces the number of iterations required to achieve a certain precision.
\end{enumerate}

\subsection{\label{ssec:impl}Implementation}

For the implementation we consider finite-length sequences in $\mathbb{C}^L,\ L \in \NN$. For the extension of the results in Appendix~\ref{sec:nonuniformFB} to finite-length sequences we refer to~\cite{Fickus:2013a}. We provide code for performing an AUDlet analysis/synthesis as part of the Matlab/Octave ``LTFAT'' toolbox \cite{ltfatnote022} available at \url{http://ltfat.sourceforge.net/}. The analysis filters are generated by the function \texttt{audfilters}. The function allows to construct at will uniform or non-uniform AUDlet FBs with integer or rational downsampling factors,\footnote{Although the results stated in Appendix~\ref{sec:nonuniformFB} are valid only for $d_{k}$'s $\in \ZZ$, rational downsampling factors can be achieved in the time domain by properly combining upsamplers and downsamplers (e.g.~\cite{Kovacevic:1993a}). In LTFAT the sampling rate changes are directly performed in the frequency domain by periodizing and folding the $Y_k(z)$'s, then performing an inverse $\mathrm{DFT}$~ \cite{
Schoerkhuber:2014a}. This technique allows to achieve rational downsampling factors at low computational costs.} thus offering flexibility in FB design. The desired number of channels can be set by specifying 
either $K$ or $V$. Using the block-processing framework proposed in \cite{Holighaus:2013a}, a real-time AUDlet analysis is also possible in LTFAT with block-stream processing \cite{ltfatnote030}. As for the prototype window $w$, the function \texttt{audfilters} uses by default a Hann window, but any FIR window can be chosen. In Sec.~\ref{sec:simu} we present results obtained with different window types. The synthesis FB is computed by the function \texttt{filterbankdual} for the cases~(i) and~(ii) mentioned above. For case~(iii), the pseudo code is presented on the Web page associated with~\cite{Necciari:2013a}. Analysis and synthesis are finally performed by the functions \texttt{filterbank} and \texttt{ifilterbank}, respectively.

\section{\label{sec:simu} Experiments}

To illustrate the properties and signal processing capabilities of the AUDlet FB, we present in this section the results from three experiments. The first experiment is a direct comparison between the proposed framework and a classic linear auditory FB, namely the gammatone FB, in terms of FB response, reconstruction error, aliasing suppression, and signal representation. The effect of the prototype filter's shape $w$ on the aliasing suppression property of the FB is also investigated. The two follow-up experiments are exemplar applications of the AUDlet FB to audio signal processing, namely a source separation and a speech de-noising experiment. For demonstration purposes all FBs were adapted to the ERB scale. The resulting AUDlet FB is thus called ``ERBlet FB'' in the following. Results on the performance of an ERBlet iterative reconstruction using CG can be found in~\cite{Necciari:2013a}. 

\subsection{Filter Bank Settings}\label{ssec:settings}

All experiments presented below feature non-uniform ERBlet and gammatone FBs. The discrete-time impulse responses of the gammatone filters were calculated by sampling and windowing the complex continuous-time gammatone IIR
\begin{equation}\label{eq:gtIR}
	h_{\mathrm{gt,k}}(t) = \alpha_{k} t^{\gamma-1} e^{2 \pi t (i\xi_k -\lambda_k)} \quad t \geq 0, k \in \{0,\ldots, K\}
\end{equation}
where $\xi$ is the filter center frequency, $\gamma$ is the gammatone filter order, $\lambda_k = \beta \, \mathrm{ERB}(\xi_k)$ determines the filter bandwidth and $\alpha_k$ is a normalization factor that constraints all filters to have the same energy. We chose $\gamma$ = 4 and $\beta$ = 1.019 to obtain a gammatone FB adapted to human auditory perception \cite{Patterson:1992a}. This FIR filter design allows for straightforward implementation but it requires rather long impulse responses to correctly approximate the filter responses at low frequencies. We used a length of 6000~samples. The gammatone synthesis FB consists of the synthesis filters $g_{\mathrm{gt,k}}[n] = \overline{h_{\mathrm{gt,k}}}[-n]$, where the bar denotes the complex conjugate, and upsampling factors $d_k$'s as the analysis FB. Using time-reversed versions of the analysis filters for synthesis might not be the best setting in terms of reconstruction error (this is discussed below) but this is the most common use of gammatone FBs in audio 
applications (e.g. \cite{Lin:2001b,Hohmann:2002a,Strahl:2009a,Gunawan:2010a,Zhao:2012a}). We therefore stick to this setting. Additionally, note that in contrast to previous band-limited gammatone FB designs, our gammatone FB covers the full range of frequencies (i.e. from 0 to the Nyquist frequency).

To achieve a fair comparison between the ERBlet and gammatone systems, both FBs were configured identically, specifically with the same spectral and temporal resolutions. In Experiment~1, a spectral resolution of $V$ = 1~filter/ERB was chosen to show results with a small density of filters. All filters were 1-ERB wide. In Experiments~2 and~3, a spectral resolution of $V = 6$ filters/ERB was chosen to achieve good signal processing performance. Since applications often require a finer resolution than that provided by 1-ERB-wide filters, especially at high frequencies where the ERBs are large, we set the bandwidths of the ERBlet and gammatone filters to one sixth of an ERB (i.e. $\Gamma_k = \mathrm{ERB}(\xi_k)/6$ and $\beta$ = 1.019/6) in those experiments. In this setting the FBs are only partly perceptually motivated, though. All ERBlet calculations were performed using a Hann window as $w$, unless otherwise stated. A set of integer downsampling factors was generated that satisfies Thm~\ref{thm:painless} for 
the Hann ERBlet. Let $R_{i}$ denote the redundancy of the resulting painless system. We then evaluated the FB performances for four redundancies $R = redfac \times R_{i}$ with $redfac$ = 0.38, 1/2, 1.0 and 2.0. For $redfac$ = 0.38 and 1/2, Thm~\ref{thm:painless} is violated and the ERBlet synthesis is done using CG.

\subsection{\label{ssec:erbvsgfb}ERBlet vs. Gammatone FB}

\begin{figure}[!t]
	\centering
	\includegraphics[width=0.9\columnwidth]{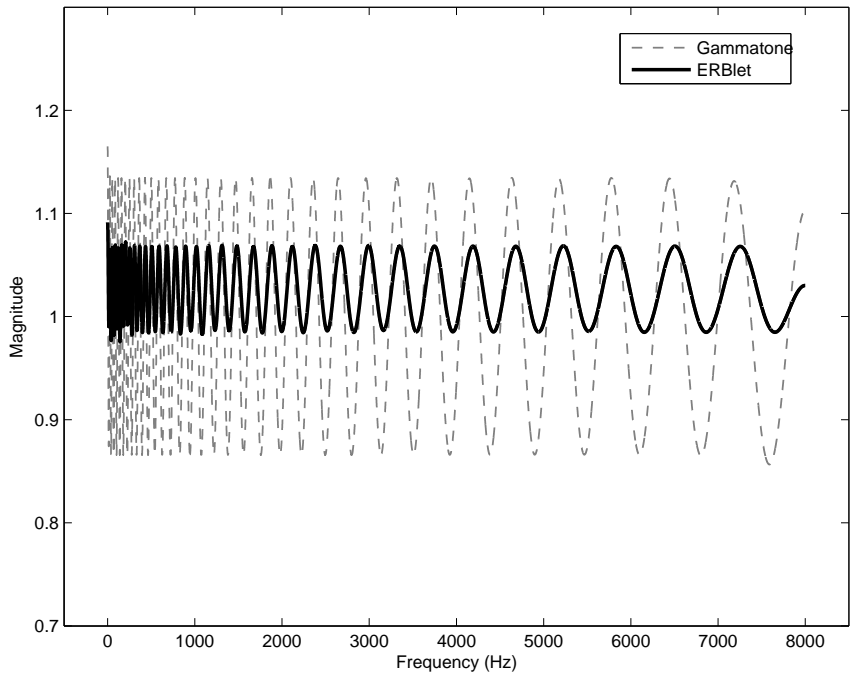}
	\caption{Magnitudes of the FB responses in the frequency range $[0; \xi_s/2]$ for the gammatone (dashed line) and ERBlet (solid line) FBs using $V$ = 1 ($K$ = 34).}
	\label{fig:erbvsgfb_resp}
\end{figure}

Fig.~\ref{fig:erbvsgfb_resp} shows the magnitudes of the ERBlet (solid line) and gammatone (dashed line) FB responses in the frequency range from zero to the Nyquist frequency. While the ERBlet FB response is rather flat across all the passband, the gammatone FB response features significant ripples. These ripples are likely to affect the reconstruction property of the gammatone FB. As is easily seen, defining synthesis filters that are time-reversed versions of the analysis filters infers that the FB response is constant over the full frequency range, cp. ~\eqref{eq:candualPL} in the appendix, noting that $\mathcal{H}_0(\xi)$ corresponds to the FB response. Therefore, the results in Fig.~\ref{fig:erbvsgfb_resp} indicate that the gammatone FB in this particular setting is not a perfect reconstruction system. Accordingly, the relative reconstruction errors for the gammatone and ERBlet FBs for the four redundancy factors are listed in Table~\ref{tab:erbvsgfb_err}. While the ERBlet scheme using the proposed 
methods always achieves perfect reconstruction up to numerical precision, using gammatone filters in the analysis and (time-reversed) in the synthesis step generates a relative error of about $10^{-1}$. A similar reconstruction error was reported in \cite{Strahl:2009a} using FIR gammatone filters and about 1~filter per ERB. Noteworthy, the present gammatone reconstructions for the two smallest redundancy factors featured audible distortions that are likely due to the ripples in the gammatone FB response and the gammatone filters' weak aliasing suppression.

\begin{figure}[!t]
	\centering
	\includegraphics[width=0.9\columnwidth]{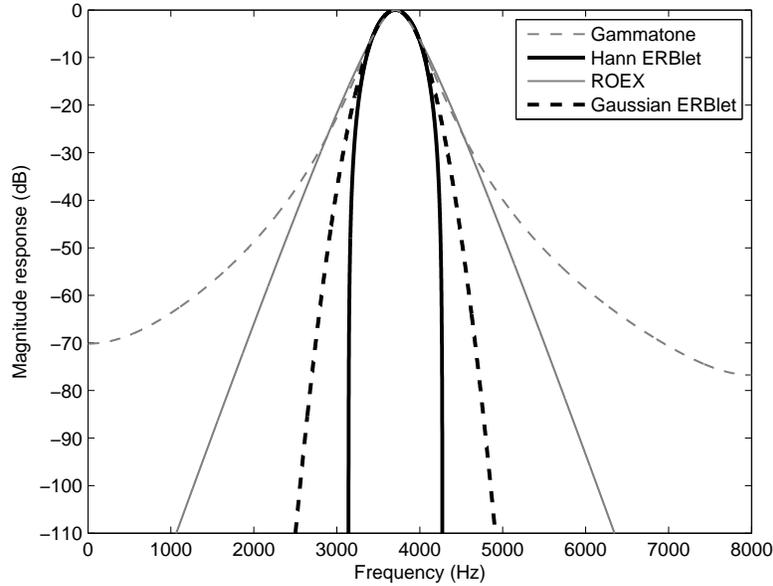}
	\caption{Magnitude responses (in dB) of $H_{\mathrm{gt,28}}(\xi)$ (dashed gray) and $H_{28}(\xi)$ (solid black) in the frequency range $[0; \xi_s/2]$. Additionally, a roex (solid gray) and a Gaussian filter (dashed black) with the same center frequency and ERB are shown.}
	\label{fig:erbvsgfb_attdB}
\end{figure}

To assess the aliasing suppression capability of each FB, the magnitude responses (in dB) of the gammatone (gray dashed) and Hann ERBlet filters (black solid) for channel $k$ = 28 are plotted in Fig.~\ref{fig:erbvsgfb_attdB} in the frequency range $[0; \xi_s/2]$. The magnitude responses of a Gaussian (black dashed) and a roex ERBlet filter (gray solid) are also shown. For the roex variant, the prototype filter $w$ was a symmetric roex(p,r) defined by its frequency response \cite{Unoki:2006a}
\begin{equation}\label{eq:roexFR}
	H_{\mathrm{roex},k}(e^{2i \pi\xi}) = (1-r)(1+p_k|\xi-\xi_k|/\xi_k)e^{-p_k|\xi-\xi_k|/\xi_k}+r,
\end{equation}
$k \in \{0,\ldots, K\}$ with $r=0$.\footnote{This setting can be easily adjusted to a parallel roex filter that better represents the auditory filters' shape than a single roex(p,r) \cite{Unoki:2006a}. However, simulations showed that using a parallel roex only deteriorates the roex' aliasing suppression outside the passband. Consequently, there is no significant impact on the reconstruction error at moderate redundancies but reconstruction error increases drastically at low redundancies, similar to the gammatone FB.}  The parameter $p_k$ was tuned to obtain a bandwidth of 1~ERB. The roex filter is an analogue of the gammatone filter but defined in the frequency domain. It is thus more easily applicable to the AUDlet FB's concept than the gammatone. 

To avoid aliasing in channel $k$ the signal must be band-limited to $\xi_s/d_k$ around $\xi_k$. In the example illustrated in Fig.~\ref{fig:erbvsgfb_attdB} $d_{28}$ = 14, $\xi_{28} \approx$~3.7~kHz and $\xi_s/d_{28} \approx$ 1200~Hz, that is the filters must have a high attenuation for frequencies outside the range $[3.1; 4.3\,\mathrm{kHz}]$. While all filters have a similar attenuation for frequencies between 3.4 and 4~kHz, the attenuations of the Hann and Gaussian ERBlet filters are superior to those of the roex and gammatone filters for other frequencies. This indicates that the ERBlet FB has a better resistance to sub-channel processing than the gammatone FB. This is verified in the two follow-up experiments.

\begin{figure*}[!t]
	  \begin{center}
		 \subfloat[]{%ERBlet analysis
		 	\includegraphics[width=0.49\columnwidth]{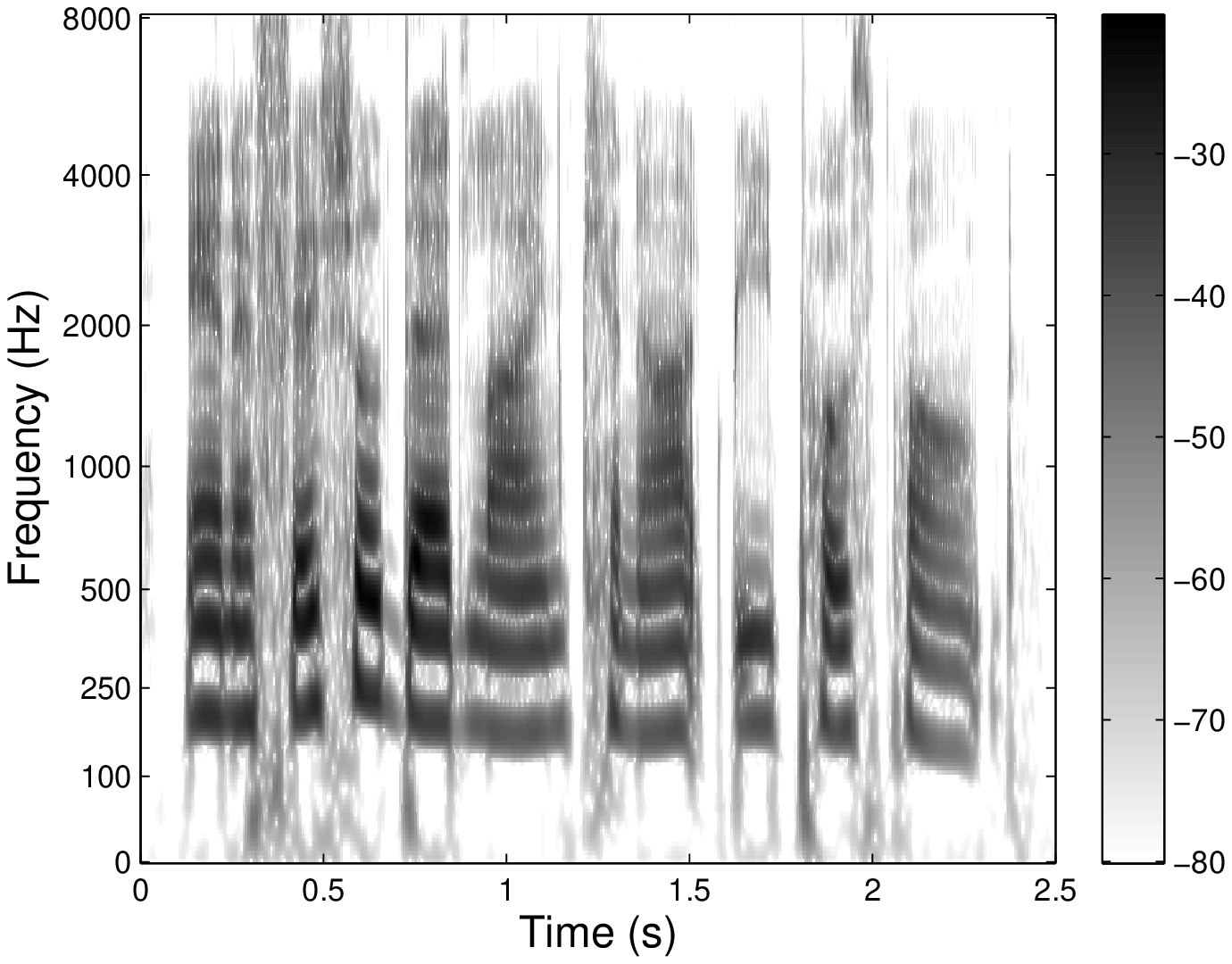}}
		 \subfloat[]{% Gammatone analysis
		 	\includegraphics[width=0.49\columnwidth]{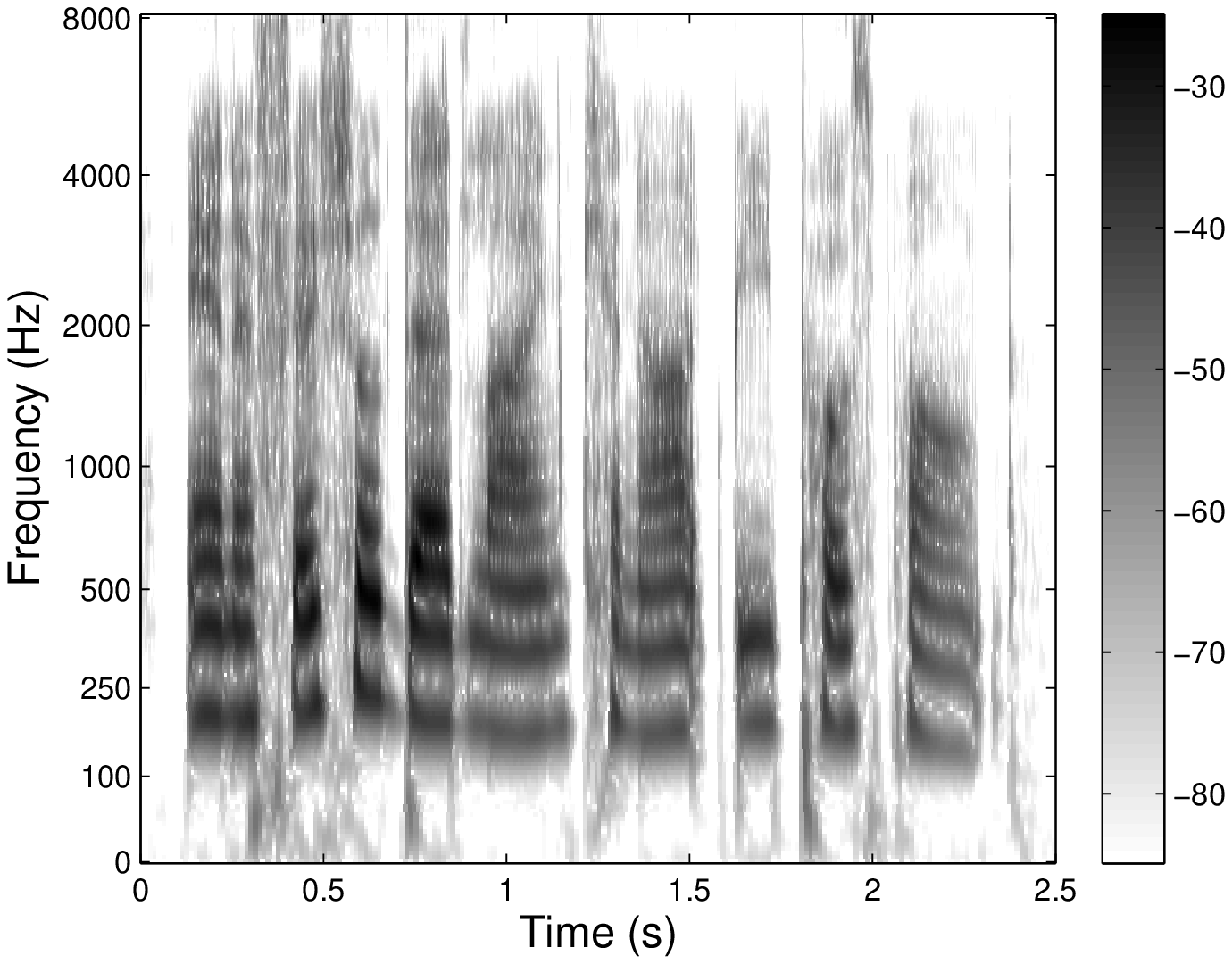}}
	  \end{center}
	  \caption{Analyses of a speech signal by (a)~the ERBlet FB and~(b)~the gammatone FB using $V$ = 6 ($K$ = 201) and $redfac$ = 2.}
	  \label{fig:erbvsgfb_img}  
	\end{figure*}

Finally, the ERBlet (a) and gammatone (b) analyses of speech signals are represented in Fig.~\ref{fig:erbvsgfb_img} for $redfac$ = 2. This experiment was performed on a female speech signal sampled at 16~kHz taken from the TIMIT database \cite{Garofolo:1993a}. To better represent the harmonics, we chose $V$ = 6. It can be seen that the two signal representations are very similar over the whole TF plane.

\begin{table}[!t]
	\caption{Relative reconstruction errors for the gammatone, ROEX and Hann ERBlet FBs using $V$ = 1 for various redundancy factors. The corresponding actual redundancies $R$'s are also indicated.}
	\label{tab:erbvsgfb_err}
	\centering
	\begin{tabular}{rcccc}
		\toprule
		$redfac$ & 0.38 & 1/2 & 1 & 2\\
		$R$ & 1.13 & 1.48 & 3.04 & 6.18\\
		\cmidrule{2-5}
		gammatone FB & 0.55 & 0.34 & 0.10 & 0.10\\
		ROEX FB & 0.67 & 0.43 & 0.12 & 0.11\\
		ERBlet FB & 1x$10^{-14}$& 4x$10^{-15}$ & 5x$10^{-16}$ & 5x$10^{-16}$\\
		\bottomrule
	\end{tabular}
\end{table}

\subsection{\label{ssec:decomp} Separation by Masking}

In this experiment, we attempt the separation of a musical source from a vocal and music mixture through a simple masking procedure using a binary mask \cite{Zhao:2012a}. We evaluate the separation using classical signal-to-noise ratio (SNR, in dB) but also the following measures proposed in~\cite{Vincent:2006a} for determining separation performance: signal-to-distortion ratio (SDR), signal-to-interference ratio (SIR) and signal-to-artifact ratio (SAR), all in dB. For their exact definition, please refer to~\cite{Vincent:2006a}. 

\begin{figure*}[!t]
	  	\begin{center}
		  \subfloat[Signal mixture]{
		  \includegraphics[width=0.45\columnwidth]{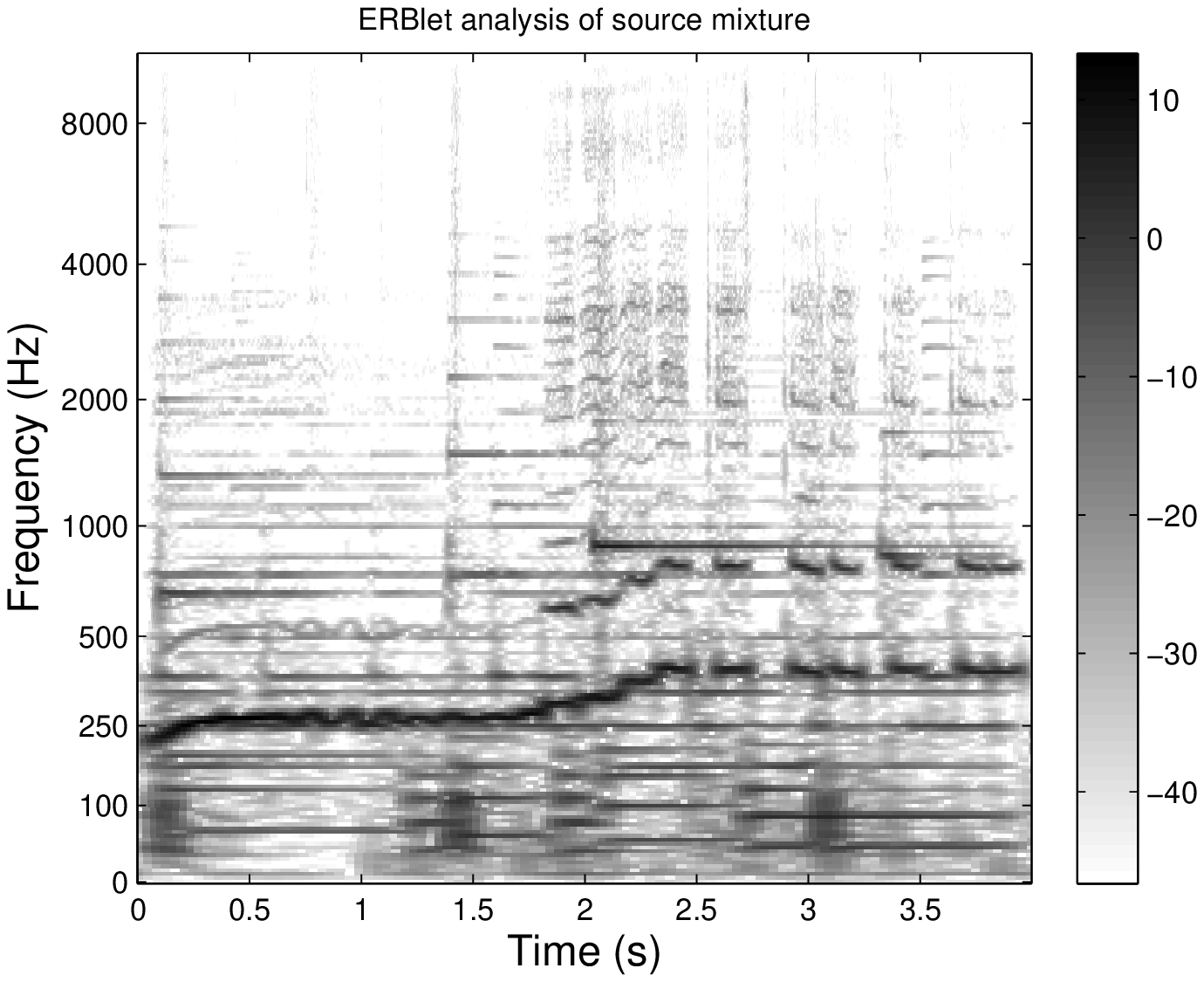}}
		  \subfloat[Ground truth signal]{ 
		   \includegraphics[width=0.45\columnwidth]{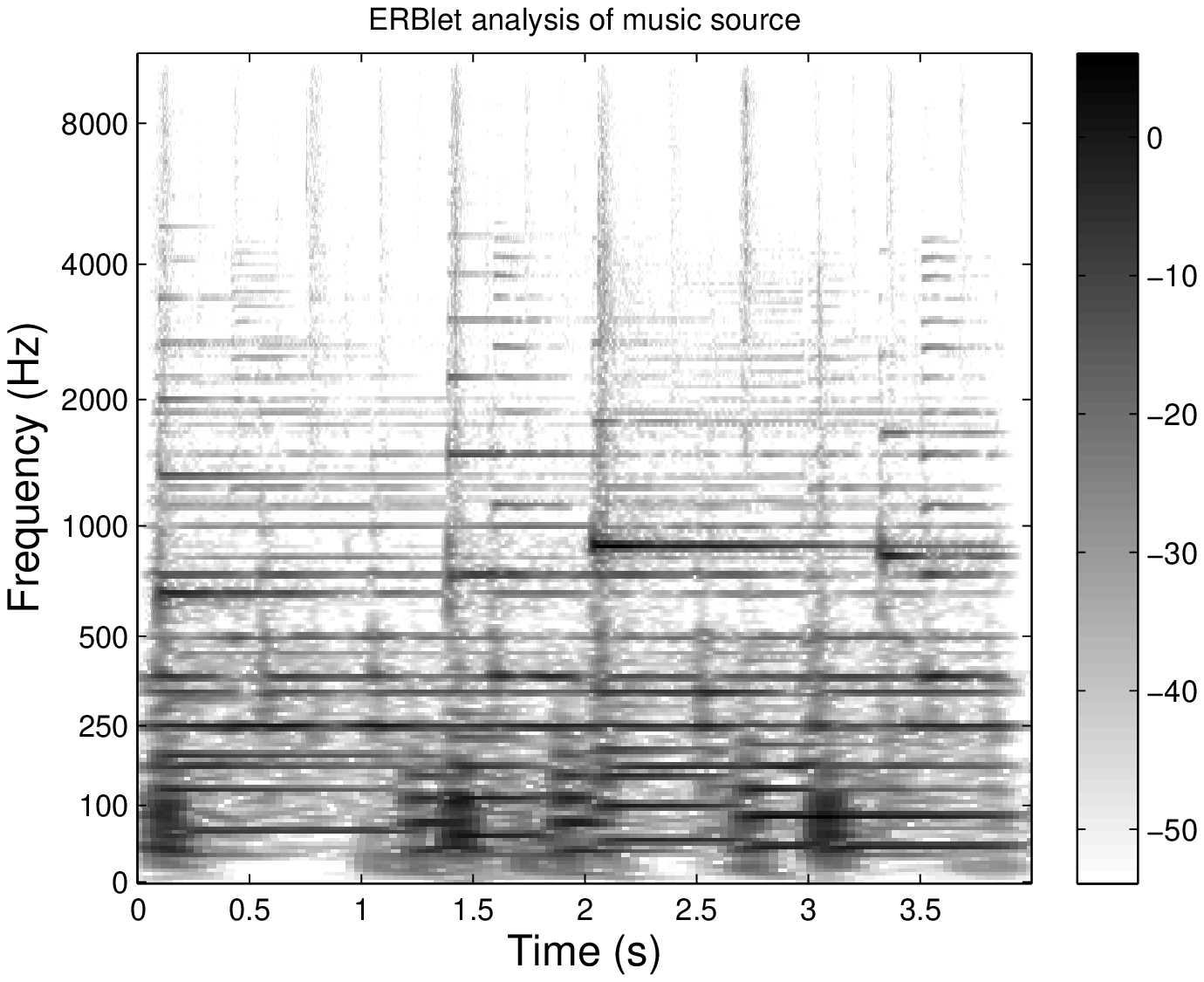}}\\
		   \subfloat[ERBlet separated signal]{
		   \includegraphics[width=0.45\columnwidth]{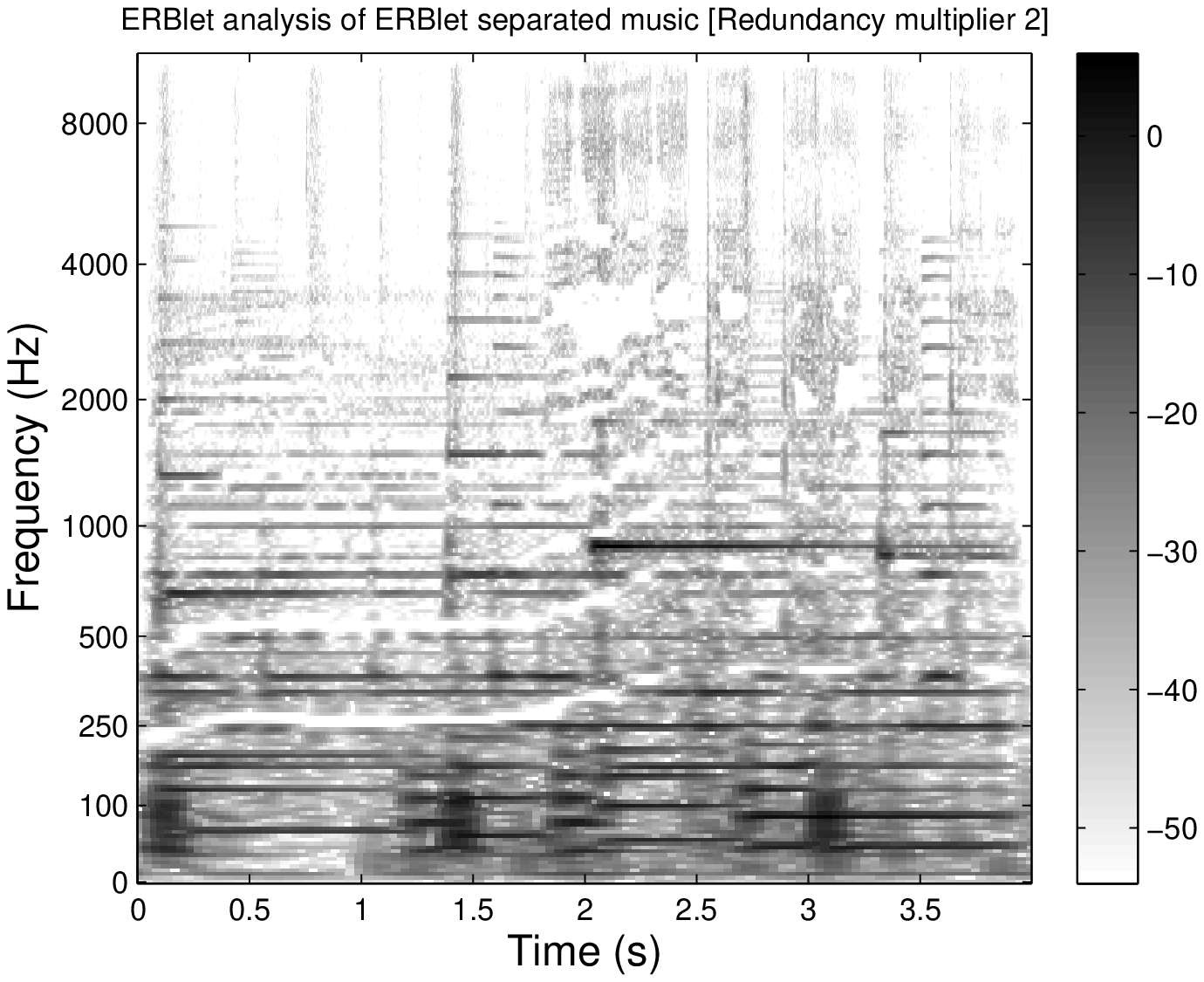}}
		   \subfloat[Gammatone separated signal]{
		   \includegraphics[width=0.45\columnwidth]{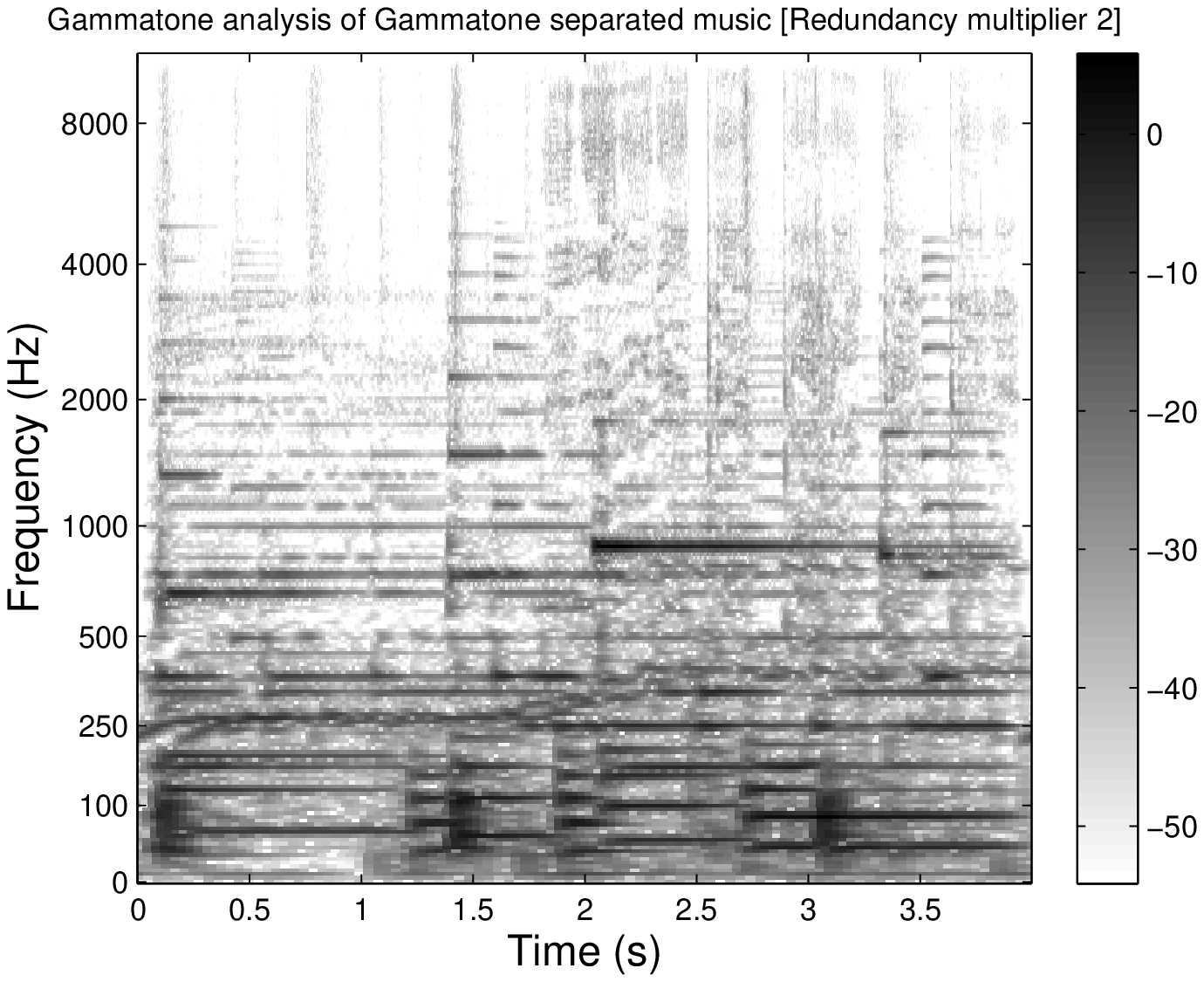}}\\
		   \subfloat[Binary separation mask]{
		   \includegraphics[width=0.45\columnwidth]{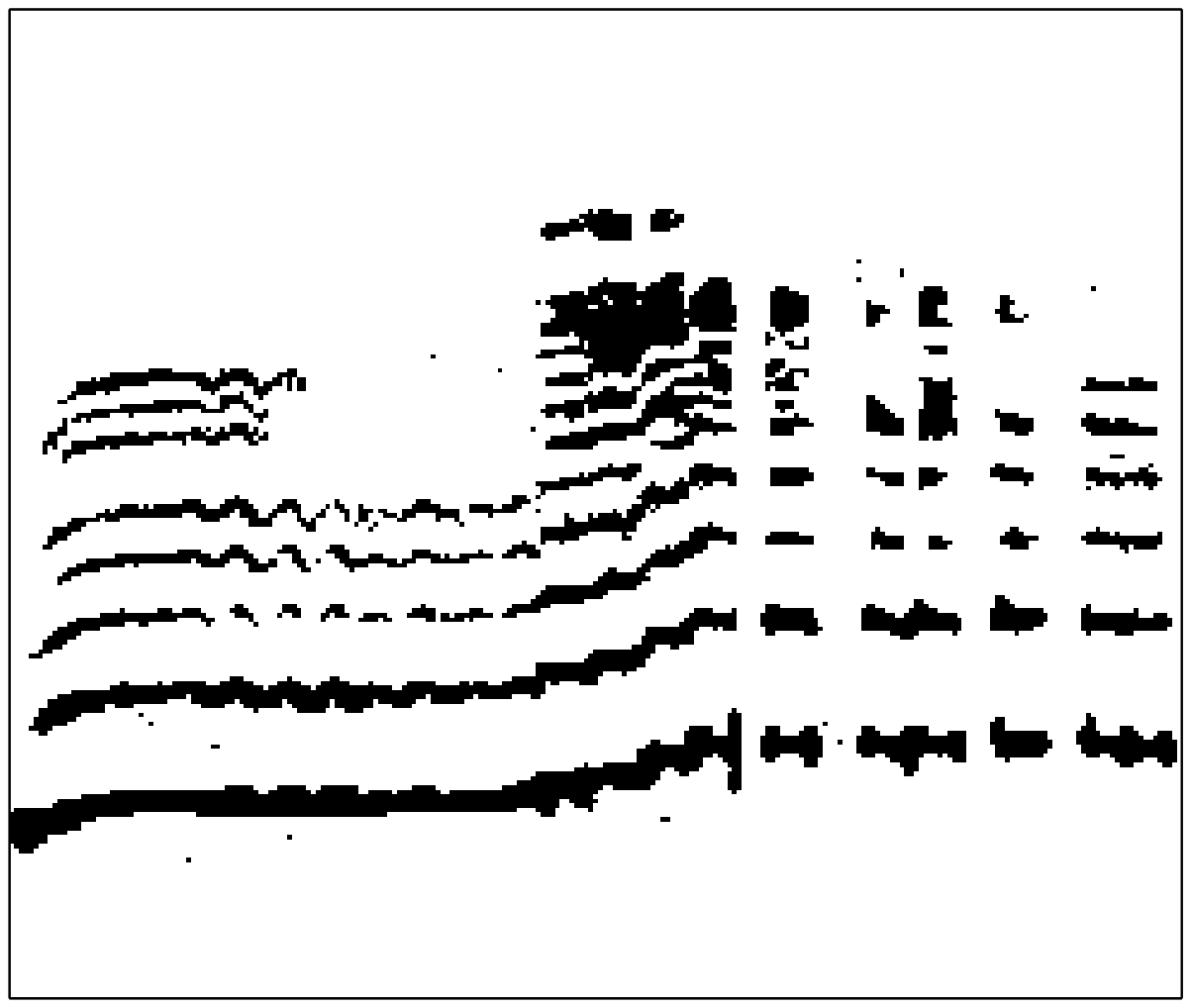}}
	 	 \end{center}
	  	\caption{Inputs (a-b) and separation results for (c)~the ERBlet and~(d)~the gammatone FB. Note how the gammatone separation shows considerably more residual energy from the vocal source than the ERBlet separation, particularly in the first section of the signal. This is a consequence of the gammatone filters' inferior TF concentration. (e)~Binary mask used in the separation process, where black pixels represent the masks $0$ entries.}\label{fig:separation_exp}  
	\end{figure*}

The binary mask was created by a combination of thresholding the ERBlet spectrogram of the vocal source and manual editing in an image processing software. Separation is performed by (i) analyzing the mixture with an ERBlet (gammatone) FB, (ii) applying the mask to the coefficients by point-wise multiplication and (iii) synthesizing with the dual FB (the time-reversed analysis FB).\footnote{Such an approach is  often used, for example, in computational auditory scene analysis \cite{Zhao:2012a} and is known in mathematical signal processing as a frame multiplier \cite{balsto09new}.} Fig.~\ref{fig:separation_exp} shows the mixture (a), ground truth (b), separated signals (c,d), and the separation mask (e) in a high redundancy setup ($redfac$ = 2). For subplots (c) and (d), the separated signal was re-analyzed using the analysis FB used in the processing step.

The separation results for low, medium and high redundancy setups are listed in Tables~\ref{tab:seplowred}--\ref{tab:sephigred}, respectively. It can be seen that the ERBlet FB slightly outperforms the gammatone FB in practically every measure, with the possible exception of SIR on the vocal source, where the results are still roughly equivalent. Note that the vocal source is not the target signal and the mask was designed to remove the voice from the music, i.e. interference in the separated voice signal is preferred over interference in the separated music source. One can also see that the performance differences between gammatone and ERBlet FBs increase with decreasing redundancy, further illustrating the superior processing stability and reconstruction quality of the ERBlet FB, in particular at low-redundancy setups.

\begin{table}[!t] 
	\begin{center}
		\caption{Quality measures for the separation experiment with low redundancy ($redfac = .38, R = 1.06$)}
		\label{tab:seplowred}	
		\begin{tabular}{ccccc} 
			 \toprule
			 & SDR & SIR & SAR & SNR \\ 
			 \cmidrule{2-5} 
			 ERBlet separated voice & $ 11.51 $ & $ 16.35 $ & $ 13.33 $ & $ 11.43 $ \\ 
			 ERBlet separated music & $  7.16 $ & $ 16.85 $ & $  7.75 $ & $  6.53 $ \\ 
			 Gammatone sep. voice & $  3.87 $ & $ 15.57 $ & $  4.29 $ & $  3.73 $ \\ 
			 Gammatone sep. music & $ -1.65 $ & $ 12.44 $ & $ -1.24 $ & $ -1.17 $ \\ 
			 \bottomrule
 		\end{tabular}
 	\end{center} 
\end{table} 

\begin{table}[!t] 
	\begin{center}
		\caption{Quality measures for the separation experiment with medium redundancy ($redfac = 1, R = 2.78$)}
		\label{tab:sepmedred}
		\begin{tabular}{ccccc} 
			 \toprule 
			 & SDR & SIR & SAR & SNR \\ 
			 \cmidrule{2-5} 
			 ERBlet separated voice & $  13.85 $ & $  16.56 $ & $  17.28 $ & $  13.65 $ \\ 
			 ERBlet separated music & $  10.92 $ & $  21.16 $ & $  11.39 $ & $   8.75 $ \\ 
			 Gammatone sep. voice & $  12.73 $ & $  16.61 $ & $  15.11 $ & $  12.55 $ \\ 
			 Gammatone sep. music & $   8.88 $ & $  20.44 $ & $   9.24 $ & $   7.65 $ \\ 
			 \bottomrule
 		\end{tabular}
 	\end{center}   
\end{table} 

\begin{table}[!t] 
	\begin{center}
		\caption{Quality measures for the separation experiment with high redundancy ($redfac = 2, R = 5.59$)}
		\label{tab:sephigred}
		\begin{tabular}{ccccc} 
			 \toprule
			 & SDR & SIR & SAR & SNR \\ 
			 \cmidrule{2-5} 
			 ERBlet separated voice & $  13.89 $ & $  16.66 $ & $  17.25 $ & $  13.69 $ \\ 
			 ERBlet separated music & $  10.89 $ & $  21.10 $ & $  11.36 $ & $   8.79 $ \\ 
			 Gammatone sep. voice & $  12.88 $ & $  16.65 $ & $  15.33 $ & $  12.67 $ \\ 
			 Gammatone sep. music & $   9.07 $ & $  20.41 $ & $   9.44 $ & $   7.77 $ \\ 
			 \bottomrule
		\end{tabular}
	\end{center} 
\end{table} 

\subsection{Speech De-Noising}

In this third and last experiment, we perform a de-noising task in the transform domain, compare to \cite{badokowto13}. 
Specifically, we apply channel-wise the soft-thresholding function \cite{Donoho:1995a}
\begin{equation*}
	y_{k_{\mathrm{thr}}} = \mathrm{sgn}(y_k)(|y_k|-\eta)_+ 
\end{equation*}
to the sub-band components $y_k[n]$ where $\eta$ is the threshold value. The de-noised signal is then obtained by applying the synthesis FB to the modified components $y_{k_{\mathrm{thr}}}$. The audio material consisted of a male and a female extract taken from the TIMIT database ($\xi_s$ = 16~kHz) \cite{Garofolo:1993a}. The signals were corrupted by Gaussian white noises of different powers. Denote by $\sigma$ the standard deviation (power) of the corrupting noise. We set the threshold parameter $\eta = \sigma$. The de-noising performance is evaluated by two measures: the SNR and segmental SNR (segSNR). The segSNR measures were computed as in \cite{Hansen:1998a} using 32-ms frames and limited to the range of -10 to 35~dB. Table~\ref{tab:denoiseres} compares the SNR (top) and segSNR (bottom) of the ERBlet and gammatone FBs for various noise powers and redundancy factors. It can be seen that the ERBlet systematically achieves higher SNR and segSNR than the gammatone. At low redundancy, the ERBlet improves the 
SNR by up to 9.5~dB (average gain = 5~dB) and the segSNR by up to 5~dB (average gain = 2.7~dB). However, the differences in SNR and segSNR are very small ($\leq$1~dB) at medium and high redundancies. Noteworthy, running this experiment without sub-sampling (i.e. $d_k$ = 1 $\forall k$) resulted in the same SNR and segSNR values as with $redfac$ = 2 for both FBs, that is it did not improve performance. Overall, the better performance of the ERBlet again illustrates its perfect reconstruction and good resistance to sub-channel processing as compared to the gammatone FB, especially at low redundancy.

\begin{table}[!t] 
	\begin{center}
		\caption{Comparison of the SNR (top) and segSNR values (bottom) of the ERBlet and gammatone FBs for two types of speech signals with different noise powers. For each signal, the corrupting noise power is indicated as the input SNR in dB.}
		\label{tab:denoiseres}
		\begin{tabular}{ccccccc} 
			 \toprule
			 Signal & \multicolumn{2}{c}{$redfac$ = .38} & \multicolumn{2}{c}{$redfac$ = 1} & \multicolumn{2}{c}{$redfac$ = 2}\\
			 \& input SNR & \multicolumn{2}{c}{$R$ = 1.1} & \multicolumn{2}{c}{$R$ = 2.9} & \multicolumn{2}{c}{$R$ = 5.9}\\			 
			 \cmidrule(lr){2-3}\cmidrule(lr){4-5}\cmidrule(lr){6-7}
			 & GFB & ERBlet & GFB & ERBlet & GFB & ERBlet\\\midrule
			 Male -5~dB & 1.64 & 3.33 & 4.21 & 4.29 & 4.24 & 4.29\\
			 Male 0~dB & 3.73 & 7.28 & 7.91 & 8.10 & 7.94 & 8.11\\
			 Male 10~dB & 5.13 & 14.67 & 14.12 & 15.35 & 14.22 & 15.35\\
			 Female -5~dB & 1.47 & 3.26 & 4.27 & 4.33 & 4.31 & 4.33\\
			 Female 0~dB & 3.49 & 7.18 & 7.95 & 8.14 & 8.00 & 8.14 \\
			 Female 10~dB & 4.51 & 14.31 & 13.95 & 15.11 & 14.08 & 15.12\\
			 \bottomrule
		\end{tabular}\\[10pt]
		\begin{tabular}{ccccccc} 
			 \toprule
			 Signal & \multicolumn{2}{c}{$redfac$ = .38} & \multicolumn{2}{c}{$redfac$ = 1} & \multicolumn{2}{c}{$redfac$ = 2}\\
			 \cmidrule(lr){2-3}\cmidrule(lr){4-5}\cmidrule(lr){6-7}
			 \& input SNR & GFB & ERBlet & GFB & ERBlet & GFB & ERBlet\\\midrule
			 Male -5~dB & -3.55 & -2.50 & -1.92 & -1.85 & -1.90 & -1.85\\
			 Male 0~dB & -1.84 & 0.22 & 0.69 & 0.86 & 0.71 & 0.86\\
			 Male 10~dB & 0.14 & 5.71 & 5.23 & 6.26 & 5.29 & 6.26\\
		 	 Female -5~dB & -4.18 & -3.25 & -2.65 & -2.61 & -2.63 & -2.61\\
			 Female 0~dB & -2.58 & -0.79 & -0.27 & -0.13 & -0.24 & -0.13\\
			 Female 10~dB & -1.02 & 3.93 & 3.79 & 4.55 & 3.88 & 4.56\\
			 \bottomrule
		\end{tabular}
	\end{center} 
\end{table}

\section{\label{sec:concl}Summary and Concluding Remarks}

The construction of an oversampled perfect reconstruction FB with filters distributed on a perceptual frequency scale has been presented. The resulting perceptually motivated FB is named ``AUDlet FB''. The FB design is directly performed in the frequency domain and allows for various filter shapes, uniform or non-uniform setting, and large downsampling factors. For redundancies $\geq$~3 (i.e. ensuring a painless system), the synthesis (dual) filters are explicitly computed. For lower redundancies, an iterative algorithm is used to compute the action of the dual FB. The TF resolution and redundancy of the FB are adaptable without affecting its perfect reconstruction property down to redundancies close to 1. Overall, the proposed system provides a simple and efficient FB design that is highly suitable for audio applications that require an analysis-synthesis framework. We provide an implementation of the AUDlet FB in the free Matlab/Octave toolbox LTFAT.

An experiment compared the AUDlet to a linear auditory FB that is widely used in audio applications, namely the gammatone FB. The results showed the better performance of the AUDlet FB with respect to the gammatone FB in terms of reconstruction error and resistance to sub-channel signal processing, especially at low redundancies. Two additional experiments demonstrated the utility of the AUDlet FB as an audio processing tool.

The proposed concept is a linear FB and, as such, does not constitute a realistic auditory filter's model, as proposed for instance in \cite{Irino:2006b,Lopez-Poveda:2001a}. In particular, we do not consider nonlinearities due to varying sound pressure levels (SPLs). Both of the cited approaches (respectively the dual-resonance nonlinear \cite{Lopez-Poveda:2001a} and dynamic compressive gammachirp FBs \cite{Irino:2006b}) feature a \textit{linear} filter in the first stage and the nonlinearities are added subsequently. It is thus conceivable that a similar nonlinear FB construction be achieved using an AUDlet FB, for instance by adding a compressive nonlinearity subsequent to the AUDlet filters. Nevertheless, this is likely to alter the stability and perfect reconstruction property of the analysis-synthesis system, especially if sub-channel processing is performed. Considering that in many applications the SPL is unknown in the signal chain (the SPL actually depends on the final listening volume), using level-
\textit{in}dependent filters is the most conservative course of action and may suffice in most cases.

To further reduce the redundancy of the AUDlet representation and improve its perceptual relevance, future work includes introducing perceptual sparsity in the transform domain. Specifically, based on the perceptual irrelevance filter proposed in \cite{Balazs:2010a} and recent data on auditory TF masking \cite{Necciari:2010a}, a binary mask will be computed and applied to the sub-channel coefficients in order to re-synthesize only the audible TF components. Furthermore, future work will focus on how to combine the AUDlet FB and knowledge of TF masking to possibly improve audio codecs. For a first approach on how to adapt the ERBlet FB for audio coding see \cite{dernecxxl15}.

\section*{Acknowledgment}

The authors would like to thank Dami\'{a}n Marelli for insightful discussions and help on the theoretical development on non-uniform FBs.

\appendix
\section{\label{sec:nonuniformFB}Invertibility of non-uniform FBs}

Consider a non-uniform FB structure with downsampling and upsampling factors $d_k \in \ZZ$ as depicted in Fig.~\ref{sfig:nonuniformFB}. In this appendix, we denote by $W_{N} = e^{2 i \pi /N}$ the $N$th root of unity.

\subsection{Perfect Reconstruction Conditions}

\begin{figure}
	\centering
	\subfloat[]{%
		%\psset{xunit=.65,yunit=.65}
\psset{unit=6mm}
\begin{pspicture}(-7,-5)(7,1)%[showgrid]
  \footnotesize
  \psset{style=RoundCorners}%,gratioWh=1.1
%  \pssignal(-6.5,0){x}{$x[n]$}
%  \pssignal(7,0){xr}{$\tilde{x}[n]$}
  \pnode(-7,0){x}
  \pnode(7,0){xr}
  
  %--- PLACING BLOCKS
  %--- 0-th channel ---
  \dotnode(-6,0){dotx0}
  \psfblock[framesize=2 1](-4.25,0){h0}{$H_{0}(z)$}
  \psdsampler[framesize=1.3 1](-2.25,0){ds0}{d_{0}}
  \dotnode(-1,0){dotY01}
  \pssignal(0,0){y0}{$y_{0}[n]$}
  \dotnode(1,0){dotY02}
  \psusampler[framesize=1.3 1](2.25,0){us0}{d_{0}}
  \psfblock[framesize=2 1](4.25,0){g0}{$G_{0}(z)$}
  \pscircleop(6.25,0){oplus0}
  
  % --- 1st channel ---
  \dotnode(-6,-1.5){dotx1}
  \psfblock[framesize=2 1](-4.25,-1.5){h1}{$H_1(z)$}
  \psdsampler[framesize=1.3 1](-2.25,-1.5){ds1}{d_1}
  \dotnode(-1,-1.5){dotY11}
  \pssignal(0,-1.5){y1}{$y_1[n]$}
  \dotnode(1,-1.5){dotY12}
  \psusampler[framesize=1.3 1](2.25,-1.5){us1}{d_1}
  \psfblock[framesize=2 1](4.25,-1.5){g1}{$G_1(z)$}
  \pscircleop(6.25,-1.5){oplus1}  
  
  %--- Placing dots ---
  \ldotsnode{90}(-6,-2.5){dots1}
  \ldotsnode{90}(-4.25,-2.5){dots2}
  \ldotsnode{90}(-2.25,-2.5){dots3}
  \ldotsnode{90}(0,-2.5){dots4}
  \ldotsnode{90}(2.25,-2.5){dots5}
  \ldotsnode{90}(4.25,-2.5){dots6}
  \ldotsnode{90}(6.25,-2.5){dots7}

    % --- Kth channel ---
  \psfblock[framesize=2 1](-4.25,-3.5){hK}{$H_{K}(z)$}
  \psdsampler[framesize=1.3 1](-2.25,-3.5){dsK}{d_K}
  \dotnode(-1,-3.5){dotYK1}
  \pssignal(0,-3.5){yK}{$y_K[n]$}
  \dotnode(1,-3.5){dotYK2}
  \psusampler[framesize=1.3 1](2.25,-3.5){usK}{d_K}
  \psfblock[framesize=2 1](4.25,-3.5){gK}{$G_K(z)$}

  %--- Connecting blocks ---
  % Input signal  
  \ncline{->}{x}{dotx0}
  \naput{$x[n]$}
  % Output signal
  \ncline{->}{oplus0}{xr}
  \naput{$\tilde{x}[n]$}  
  
  \nclist{ncline}{x,h0,ds0,dotY01}
  \nclist{ncline}{dotY02,us0,g0,oplus0,xr}
  \nclist{ncline}{dotx1,h1,ds1,dotY11}
  \nclist{ncline}{dotY12,us1,g1,oplus1}
  \ncline{dotx0}{dotx1}
  \ncline{dotx1}{dots1}
  \ncline{oplus0}{oplus1}
  \ncline{oplus1}{dots7}
  \nclist{ncline}{hK,dsK,dotYK1}
  \nclist{ncline}{dotYK2,usK,gK}
  \ncangle[angleA=-90,angleB=-180,armB=0]{dots1}{hK}
  \ncangle[angleB=-90,armB=0]{gK}{dots7}%
\end{pspicture}%
	\label{sfig:nonuniformFB}}\\
	\vspace{-16pt}
	\subfloat[]{%		
		%\psset{xunit=.65,yunit=.65}
\psset{unit=6mm}
\begin{pspicture}(-7,-10.9)(7,1)%[showgrid]
  \footnotesize
  \psset{style=RoundCorners}%,gratioWh=1.1
  %\pssignal(-6.5,0){x}{$x[n]$}
  \pnode(-7,0){x}
  %\pssignal(7,0){xr}{$\tilde{x}[n]$}
  \pnode(7,0){xr}
  
  %--- PLACING BLOCKS
  %--- Decomposition of the 0-th channel ---
  %--- Subchannel 0.0:
  \dotnode(-6,0){dotx00}
  \psfblock[framesize=2.7 1](-4.25,0){h00}{$H_{0}^{(0)}(z)$}
  \psdsampler[framesize=1 1](-2.25,0){ds00}{D}
  \dotnode(-1.5,0){dot001}
  \pssignal(0,0){y00}{$y_{0}^{(0)}[n]$}
  \dotnode(1.5,0){dot002}
  \psusampler[framesize=1 1](2.25,0){us00}{D}
  \psfblock[framesize=2.7 1](4.25,0){g00}{$G_{0}^{(0)}(z)$}
  \pscircleop(6.25,0){oplus00}
  
%  %%--- Subchannel 0.1:
  \dotnode(-6,-1.5){dotx01}
  \psfblock[framesize=2.7 1](-4.25,-1.5){h01}{$H_{0}^{(1)}(z)$}
  \psdsampler[framesize=1 1](-2.25,-1.5){ds01}{D}
  \dotnode(-1.5,-1.5){dot011}
  \pssignal(0,-1.5){y01}{$y_{0}^{(1)}[n]$}
  \dotnode(1.5,-1.5){dot012}
  \psusampler[framesize=1 1](2.25,-1.5){us01}{D}
  \psfblock[framesize=2.7 1](4.25,-1.5){g01}{$G_{0}^{(1)}(z)$}
  \pscircleop(6.25,-1.5){oplus01}
%  
%  %--- Placing dots ---
  \ldotsnode{90}(-6,-2.5){dots01}
  \ldotsnode{90}(-4.25,-2.5){dots02}
  \ldotsnode{90}(-2.25,-2.5){dots03}
  \ldotsnode{90}(2.25,-2.5){dots04}
  \ldotsnode{90}(4.25,-2.5){dots05}
  \ldotsnode{90}(6.25,-2.5){dots06}

  %%--- Subchannel 0.k:
  \dotnode(-6,-3.5){dotx0k}
  \psfblock[framesize=2.7 1](-4.25,-3.5){h0k}{$H_{0}^{(q_0-1)}(z)$}
  \psdsampler[framesize=1 1](-2.25,-3.5){ds0k}{D}
  \dotnode(-1.5,-3.5){dot0k1}
  \pssignal(0,-3.5){y0k}{$y_{K}^{(q_0-1)}[n]$}
  \dotnode(1.5,-3.5){dot0k2}
  \psusampler[framesize=1 1](2.25,-3.5){us0k}{D}
  \psfblock[framesize=2.7 1](4.25,-3.5){g0k}{$G_0^{(q_0-1)}(z)$}
  \pscircleop(6.25,-3.5){oplus0k}
  
  %  %--- Placing more dots ---
  \ldotsnode{90}(-6,-5){dots11}
  \ldotsnode{90}(-4.25,-5){dots12}
  \ldotsnode{90}(4.25,-5){dots13}
  \ldotsnode{90}(6.25,-5){dots14}
  
  % --- Insert a text mentioning the number of channels...
  \pssignal(0,-5){kchannels}{$\sum_{k=1}^{k=K-1} q_k = \frac{D}{d_k}$ channels}
  
  %--- Subchannel K.0:
  \dotnode(-6,-6.5){dotxK0}
  \psfblock[framesize=2.7 1](-4.25,-6.5){hK0}{$H_{K}^{(0)}(z)$}
  \psdsampler[framesize=1 1](-2.25,-6.5){dsK0}{D}
  \dotnode(-1.5,-6.5){dotYK01}
  \pssignal(0,-6.5){yK0}{$y_{K}^{(0)}[n]$}
  \dotnode(1.5,-6.5){dotYK02}
  \psusampler[framesize=1 1](2.25,-6.5){usK0}{D}
  \psfblock[framesize=2.7 1](4.25,-6.5){gK0}{$G_{K}^{(0)}(z)$}
  \pscircleop(6.25,-6.5){oplusK0}
  
%  %%--- Subchannel K.1:
  \dotnode(-6,-8){dotxK1}
  \psfblock[framesize=2.7 1](-4.25,-8){hK1}{$H_{K}^{(1)}(z)$}
  \psdsampler[framesize=1 1](-2.25,-8){dsK1}{D}
  \dotnode(-1.5,-8){dotYK11}
  \pssignal(0,-8){yK1}{$y_{K}^{(1)}[n]$}
  \dotnode(1.5,-8){dotYK12}
  \psusampler[framesize=1 1](2.25,-8){usK1}{D}
  \psfblock[framesize=2.7 1](4.25,-8){gK1}{$G_{K}^{(1)}(z)$}
  \pscircleop(6.25,-8){oplusK1}
  
%  
%  %--- Placing dots ---
  \ldotsnode{90}(-6,-9){dotsK1}
  \ldotsnode{90}(-4.25,-9){dotsK2}
  \ldotsnode{90}(-2.25,-9){dotsK3}
  \ldotsnode{90}(2.25,-9){dotsK4}
  \ldotsnode{90}(4.25,-9){dotsK5}
  \ldotsnode{90}(6.25,-9){dotsK6}
  
  %%--- Subchannel K.k:

  \psfblock[framesize=2.7 1](-4.25,-10){hKk}{$H_{K}^{(q_K-1)}(z)$}
  \psdsampler[framesize=1 1](-2.25,-10){dsKk}{D}
  \dotnode(-1.5,-10){dotYKk1}
  \pssignal(0,-10){yKk}{$y_{K}^{(q_K-1)}[n]$}
  \dotnode(1.5,-10){dotYKk2}
  \psusampler[framesize=1 1](2.25,-10){usKk}{D}
  \psfblock[framesize=2.7 1](4.25,-10){gKk}{$G_{K}^{(q_K-1)}(z)$}

  %--- Connecting blocks ---
  % Input signal  
  \ncline{->}{x}{dotx00}
  \naput{$x[n]$}
  % Output signal
  \ncline{->}{oplus00}{xr}
  \naput{$\tilde{x}[n]$}
  
  % The rest...
  % Channel 00
  \nclist{ncline}{x,h00,ds00,dot001}
  \nclist{ncline}{dot002,us00,g00,oplus00}
  % Channel 01
  \nclist{ncline}{dotx01,h01,ds01,dot011}
  \nclist{ncline}{dot012,us01,g01,oplus01}
  % Channel 0k
  \nclist{ncline}{dotx0k,h0k,ds0k,dot0k1}
  \nclist{ncline}{dot0k2,us0k,g0k,oplus0k}
  % Vertical connections
  \ncline{dotx00}{dotx01}
  \ncline{dotx01}{dots01}
  \ncline{dots01}{dotx0k}
  \ncline{oplus01}{oplus00}
  
  \ncline{dotx0k}{dots11}
  \ncline{dots11}{dotxK0}
  \ncline{oplus0k}{dots14}
  \ncline{dots14}{oplusK0}
  
  % Channel K0
  \nclist{ncline}{dotxK0,hK0,dsK0,dotYK01}
  \nclist{ncline}{dotYK02,usK0,gK0,oplusK0}
  % Channel K1
  \nclist{ncline}{dotxK1,hK1,dsK1,dotYK11}
  \nclist{ncline}{dotYK12,usK1,gK1,oplusK1}
  % Channel Kk
  \nclist{ncline}{dotxKk,hKk,dsKk,dotYKk1}
  \nclist{ncline}{dotYKk2,usKk,gKk,oplusKk}
  % Vertical connections
  \ncline{dotxK0}{dotxK1}
  \ncline{dotxK1}{dotsK1}
  \ncline{oplusK1}{oplusK0}
  
  % Bottom connections
  \ncangle[angleA=-90,angleB=-180,armB=0]{dotsK1}{hKk}
  \ncangle[angleB=-90,armB=0]{gKk}{dotsK6}%
\end{pspicture}%
	\label{sfig:equniformFB}}
	\caption{(a) General structure of a non-uniform analysis-synthesis FB and~(b) an equivalent uniform FB~\cite{Akkarakaran:2003a}. The terms $H_k^{(l)}$ and $G_k^{(l)}$ in~(b) correspond to the $z$-transforms of the terms $h_{k}^{(l)}$ and $g_{k}^{(l)}$ defined in~\eqref{eq:hkl} and~\eqref{eq:gkl}, respectively.}
	\label{fig:FB_blocks}
\end{figure}
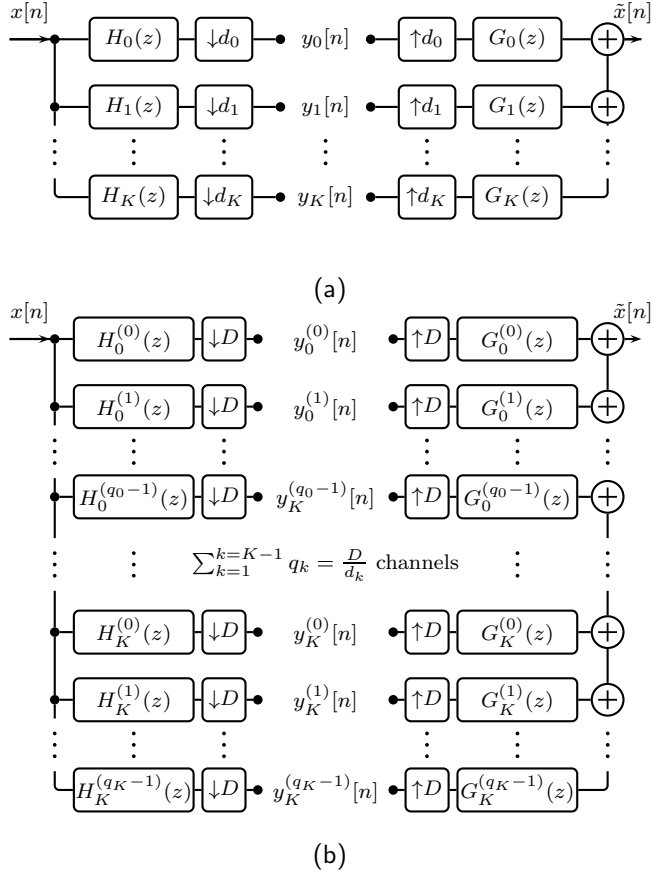

 The sub-band components of the system represented in Fig.~\ref{sfig:nonuniformFB} are given in the discrete-time domain by 
\begin{equation}\label{eq:yknonuniform}
	y_{k}[n] \, = \, \downarrow_{d_{k}}\left\{ h_{k}*x\right\}[n].
\end{equation}
The output signal is $\tilde{x}[n] \, =\, \sum_{k=0}^{K}\left(g_{k}*\uparrow_{d_{k}}\left\{y_{k}\right\} \right)[n]$. Such an analysis-synthesis system provides perfect reconstruction if $\tilde{x}[n] = x[n]$ (up to a delay factor). When $d_k = D \, \forall \, k \in \{0 \ldots K\}$ the system in Fig.~\ref{sfig:nonuniformFB} results in a uniform FB. The perfect reconstruction conditions for \textit{uniform} FBs have been largely treated in the literature (see e.g. \cite{Kovacevic:1993a,Vaidyanathan:1993a}). To treat the \textit{non-uniform} case, one possibility is to decompose the non-uniform system into a larger equivalent uniform system \cite{Kovacevic:1993a,Akkarakaran:2003a}, as shown in Fig.~\ref{sfig:equniformFB}. Denote $D = \lcm(d_k : k \in \{0,\cdots,K\})$ and $q_k = D/d_k$. Each $k$-th channel of the non-uniform system is decomposed into $q_k$ channels in the equivalent uniform system, which then features $\sum_{k=0}^{k=K} q_k$ channels in total with the downsampling factor $D$ in all channels.\
footnote{Note that for a maximally decimated non-uniform FB, i.e. when $\sum_k 1/d_k = 1$, the equivalent uniform FB features $D$ channels.} Note that the filters in Fig.~\ref{sfig:equniformFB} are various delayed versions of those in Fig.~\ref{sfig:nonuniformFB}. The sub-band components for $l \in \left\{ 0,\cdots,q_{k}-1\right\}$ in Fig.~\ref{sfig:equniformFB} can be expressed in the discrete-time domain as
\begin{eqnarray}\nonumber
	y_{k}^{(l)}[n] & = & y_{k}[n q_{k} - l]\\\label{eq:hkl}
 	& = & \downarrow_{D}\{\underbrace{h_{k}*\delta_{ld_{k}}}_{:= h_{k}^{(l)}}* x\} [n] .
\end{eqnarray}
Grouping the $q_k$ sub-band components resulting from the $k$-th sub-band yields $y_{k}[n] = \sum_{l=0}^{q_{k}-1}\uparrow_{q_{k}}\left\{y_{k}^{(l)}\right\} [n+l]$ and the output signal of the equivalent uniform FB can be written as
\begin{eqnarray}\label{eq:gkl}
	\tilde{x}[n] & = & \sum_{k=0}^{K}\sum_{l=0}^{q_{k}-1}\left(\underbrace{g_{k}*\delta_{-ld_{k}}}_{:= g_{k}^{(l)}}*\uparrow_{D}\left\{y_{k}^{(l)}\right\} \right)[n].
\end{eqnarray}

These results can be written in the frequency domain by simply taking the $z$-transform. The sub-band components \textit{after} upsampling by $D$ yield
\begin{eqnarray*}
	Y_{k}^{(l)}\left(z^{D}\right) & = & \frac{1}{D}\sum_{j=0}^{D-1} H_{k}^{(l)}\left(W_{D}^{j}z\right)X\left(W_{D}^{j}z\right)\\
 	& = & \frac{1}{D}\sum_{j=0}^{D-1}W_{D}^{-jld_{k}}z^{-ld_{k}} \underbrace{H_{k}\left(W_{D}^{j}z\right)X\left(W_{D}^{j}z\right)}_{\text{for }j \neq 0: \text{ aliasing terms}}
\end{eqnarray*}

\noindent and the output finally yields
\begin{align}\notag
	\widetilde{X}(z) & = \sum_{k=0}^{K}\sum_{l=0}^{q_{k}-1} G_{k}^{(l)}(z)Y_{k}^{(l)}\left(z^{D}\right)\\\label{eq:xtildez}
 	& = \frac{1}{D}\sum_{k=0}^{K}\sum_{j=0}^{D-1} A_{k,j} G_{k}(z) H_{k}\left(W_{D}^{j}z\right)X\left(W_{D}^{j}z\right)
\end{align}
where
\[
	A_{k,j} := \sum_{l=0}^{q_{k}-1} W_{D}^{-jld_{k}} = \sum_{l=0}^{q_{k}-1} e^{-2i\pi jl/q_{k}} = \left\{%
	\begin{array}{ll}
		q_k & \text{if $j$ is a}\\
		& \text{multiple of $q_k$}\\
		0 & \text{otherwise.}
	\end{array}\right.
\]
Equation~\eqref{eq:xtildez} can be formulated as a matrix multiplication
\begin{equation}\label{eq:polyphas1}
      \widetilde{X}(z) = \frac{1}{D}\left[X(W_{D}^{0}z) \cdots X(W_{D}^{D-1}z)\right] \bdi H(z) \bdi G(z)
\end{equation}
where $\bdi G(z) := 
 \left[G_0(z), \hdots , G_K(z) \right]^T$ and  the $D \times K$ alias cancellation matrix $\bdi H(z) = \left[\bdi h_0(z) \cdots \bdi h_K(z)\right]$, cf.~\cite{Akkarakaran:2003a}, with:
\[
	\bdi h_{k}(z) = q_k \left[%
	\begin{array}{c}
		\bdi h'_k(z)\\
	    \bdi h'_k\left(W_{D}^{q_k}z\right)\\
	    \vdots \\
	    \bdi h'_k\left(W_{D}^{(d_k-1)q_k}z\right)
	\end{array}\right]
\] and
\[
	\bdi h'_k(z) = \Big[H_k(z) \underbrace{\, 0 \, \cdots \, 0 \,}_{q_{k}-1 \text{ zeros}}\Big]^T .
\]
The perfect reconstruction condition then reduces to
\begin{equation}\label{eq:PRcondition}
	\bdi H(z) \, \bdi G(z)  = \left[D \; 0 \cdots 0 \; \right]^{T},
\end{equation}
which means that all aliasing terms have to be canceled by the synthesis filters. Equation~\eqref{eq:PRcondition} is useful to determine whether a complete FB provides perfect reconstruction. It may however fail to provide straightforward or efficient ways to find, given fixed analysis parameters $h_k$'s and $d_k$'s, fitting synthesis filters and downsampling factors, although it can sometimes be used to determine whether such a system even exists.

\subsection{\label{ssec:frametheory}Connection to Frame Theory}

Since our construction of perceptually motivated FBs emphasizes stable reconstruction from the FB coefficients, it seems worthwhile to mention that inversion of non-uniform FBs can also be investigated using frame theory, the mathematical theory of stable, redundant spanning sets of functions. For uniform FBs, this connection has been explored in depth (e.g. \cite{Bolcskei:1998a,Cvetkovic:1998a,Fickus:2013a,badokowto13}). For non-uniform FBs, the connections have, to our knowledge, not been stated explicitly in the literature although implicitly used in recent work (e.g. \cite{Holighaus:2013a},\cite{Bayram:2013a}). A frame over the space of finite energy sequences $\ell_2(\mathbb Z)$ is a (possibly redundant) family of functions spanning the space in a stable fashion, in the sense of inequality \eqref{eq:framineq1} below. The central observation linking FBs to frames is that 
\[
 y_{k}[n] =  \downarrow_{d_{k}}\left\{ h_{k}*x\right\}[n] = \langle x, \overline{h_k}[nd_k -\cdot] \rangle .
\]
\noindent Hence, the FB coefficients with respect to the filters $h_k$ and downsampling factors $d_k$ equal the frame coefficients of the system $\left(\overline{h_k}[nd_k -\cdot]\right)_{k,n}$. As a consequence, the FB allows for numerically stable perfect reconstruction if and only if $0< A\leq B < \infty$ exist such that 
\begin{equation}\label{eq:framineq1}
 A\|x\|^2 \leq \sum_k \|y_k\|^2 \leq B\|x\|^2, \text{ for all } x\in\ell^2(\mathbb{Z})
\end{equation}
where $A$ and $B$ are respectively the lower and upper frame bounds of the system.\footnote{Boundedness of $h_k$ for all $k$ is sufficient for the existence of $B$ since the number of channels is finite.}

Therefore, instead of verifying the perfect reconstruction conditions directly, we can equivalently employ techniques from frame theory to determine the inversion of the FB analysis operation and/or an appropriate synthesis system. In particular, a dual (synthesis) frame can be found by applying the inverse of the frame operator $\mathbf{S}$ defined by 
\[ 
 \mathbf{S} \, x[n] = \sum_{n,k} \langle x, \overline{h_k}[nd_k -\cdot] \rangle h_k[nd_k -\cdot],
\]
to each frame element. More precisely, we will use the inherent structure of the frequency domain variant, the matrix Fourier transform \cite{Balazs:2006a} of $\mathbf{S}$, $\widehat{\mathbf{S}} = \mathrm{DTFT}\,\mathbf{S}\,\mathrm{DTFT}^{-1}$ of the frame operator. By applying Eq. \ref{eq:polyphas1}, we can easily see, using $\mathcal{Z}(\overline{h[-\cdot]})(z) = \overline{\mathcal{Z}(h)(1/\overline{z})}$, that the action of $\bd S_{\widehat{\Phi}}$ is given by 
\begin{equation}
\begin{split}
  \widehat{\mathbf{S}}X(z) & = \frac{1}{D}\left[X(W_{D}^{0}z) \cdots X(W_{D}^{D-1}z)\right] \bdi H(z) \left[%
      	\begin{array}{c}
        	\overline{H_0(1/\overline{z})}\\
        	\vdots \\
        	\overline{H_K(1/\overline{z})}
        \end{array}\right]\\
  & = \left[Y_0(z^{d_0}) \cdots Y_K(z^{d_K})\right]\left[\overline{H_0(1/\overline{z})} \cdots \overline{H_K(1/\overline{z})}\right]^T.
\end{split}
\end{equation}
Defining 
\begin{equation}
	\left[\begin{array}{c}
      	\mathcal{H}_0(\xi)\\
       	\vdots\\
        \mathcal{H}_{D-1}(\xi)
        \end{array}\right] :=
  	\frac{1}{D}\bdi H(e^{2i\pi\xi})\left[\begin{array}{c}
      	\overline{H_0(e^{2\pi i\xi})}\\
       	\vdots\\
       	\overline{H_K(e^{2\pi i\xi})}
        \end{array}\right]
\end{equation}
for $\xi\in\mathbb{T} = \RR/\ZZ$, we obtain the following results, see \cite{Balazs:2011a} and \cite{DBLP:journals/ijwmip/DorflerM14} for the mathematical context.% {\xxl Here $\mathcal{H}_k$ are defined, right?}

\begin{Thm}\label{thm:painless}
  If, for every $0 \leq k \leq K$, the filter $h_k$ is band-limited on an interval of length $1/d_k$, i.e. there is $I_k \subseteq \mathbb{T}$ with $|I_k|\leq 1/d_k$ such that $H_k(e^{2i\pi\xi}) = 0$ for almost every $\xi\in \mathbb{T}\setminus I_k$. 
  Then $\mathcal{H}_k$ equals the zero function for $k=1,\ldots,K$ and the FB comprised of the filters $h_k$'s and downsampling factors $d_k$'s forms a frame if and only if there are $A,B$ such that
  \begin{equation}\label{eq:maindiag}
    0 < A \leq \mathcal{H}_0(\xi) \leq B < \infty, \text{ for a.e. } \xi\in \mathbb{T}.
  \end{equation}
  Moreover, a dual FB frame with upsampling factors $d_k$'s is given by the filters $g_k$'s defined by
  \begin{equation}\label{eq:candualPL}
    G_k(e^{2i\pi\xi}) = \frac{H_k(e^{2i\pi\xi})}{\mathcal{H}_0(\xi)} \text{ a.e.}
  \end{equation}
\end{Thm}

Although the implications of Theorem \ref{thm:painless} are relatively well understood, frame theory provides a simple generalization that is very useful when combined with more sophisticated inversion techniques.

\begin{Thm}\label{thm:diagdom}
  Let $h_k$'s and $d_k$'s for $0 \leq k \leq K$ define an analysis FB. If there are $0< A_0 \leq B_0 < \infty$ with 
  \begin{equation}\label{eq:diagdom1}
    A_0 \leq \mathcal{H}_0(\xi) \pm \sum_{n=1}^{D-1} |\mathcal{H}_n(\xi)| \leq B_0,\text{ for a.e. } \xi\in\mathbb{T},
  \end{equation}
  then the FB defined by $h_k$'s and $d_k$'s forms a frame.
\end{Thm}

Although reconstruction can be implemented by rewriting the FB as a uniform FB and computing the dual uniform FB, this is only feasible if $\sum_k q_k$ is not too large. However, any FB frame admits reconstruction by a \emph{conjugate gradients} (CG) algorithm~\cite{Grochenig:1993a,Trefethen:1997a} solving
\begin{equation}\label{eq:CG}
	\widehat{\mathbf{S}}X(z) = \sum_{k=0}^{K} Y_k(z^{d_{k}}) H_k(z).
\end{equation}
The number of CG steps necessary for convergence depends solely on the condition number of $\widehat{\mathbf{S}}$. Additionally, if the conditions of Thm~\ref{thm:diagdom} are satisfied and the second term in \eqref{eq:diagdom1} is small, then $\widehat{\mathbf{S}}$ is diagonal dominant and its diagonal equals $\mathcal{H}_0$. In this setting, using $\mathcal{H}_0^{-1}$ as a diagonal preconditioner has been shown to further increase convergence speed~\cite{Grochenig:1993a,Balazs:2006a,Necciari:2013a}. This method often allows for efficient inversion even if direct computation of a dual uniform FB is not feasible.

\bibliographystyle{amsplain}
\bibliography{IEEEabrv,ERBlet_jrnl_refs}

\providecommand{\bysame}{\leavevmode\hbox to3em{\hrulefill}\thinspace}
\providecommand{\MR}{\relax\ifhmode\unskip\space\fi MR }
\providecommand{\MRhref}[2]{%
  \href{http://www.ams.org/mathscinet-getitem?mr=#1}{#2}
}
\providecommand{\href}[2]{#2}
\begin{thebibliography}{10}

\bibitem{Akkarakaran:2003a}
Sony Akkarakaran and PP~Vaidyanathan, \emph{Nonuniform filter banks: new
  results and open problems}, Beyond wavelets, Studies in Computational
  Mathematics, vol.~10, Elsevier, 2003, pp.~259--301.

\bibitem{Balazs:2011a}
P.~Balazs, M.~D{\"o}rfler, N.~Holighaus, F.~Jaillet, and G.~Velasco,
  \emph{Theory, implementation and applications of nonstationary {G}abor
  frames}, J. Comput. Appl. Math. \textbf{236} (2011), no.~6, 1481--1496.

\bibitem{badokowto13}
P.~Balazs, M.~D{\"o}rfler, M.~Kowalski, and B.~Torr{\'e}sani, \emph{Adapted and
  adaptive linear time-frequency representations: a synthesis point of view},
  IEEE Signal Process. Mag. \textbf{30} (2013), no.~6, 20--31.

\bibitem{Balazs:2006a}
P.~Balazs, H.~G. Feichtinger, M.~Hampejs, and G.~Kracher, \emph{Double
  preconditioning for {G}abor frames}, IEEE Trans. Signal Process. \textbf{54}
  (2006), no.~12, 4597--4610.

\bibitem{Balazs:2010a}
P.~Balazs, B.~Laback, G.~Eckel, and W.~A. Deutsch, \emph{Time-frequency
  sparsity by removing perceptually irrelevant components using a simple model
  of simultaneous masking}, IEEE Trans. Audio, Speech, Language Process.
  \textbf{18} (2010), no.~1, 34--49.

\bibitem{Baumgarte:2002a}
F.~Baumgarte, \emph{Improved audio coding using a psychoacoustic model based on
  a cochlear filter bank}, IEEE Speech Audio Process. \textbf{10} (2002),
  no.~7, 495--503.

\bibitem{Bayram:2013a}
I~Bayram, \emph{An analytic wavelet transform with a flexible time-frequency
  covering}, IEEE Trans. Signal Process. \textbf{61} (2013), no.~5, 1131--1142.

\bibitem{Bolcskei:1998a}
Helmut B{\"o}lcskei, Franz Hlawatsch, and Hans Feichtinger,
  \emph{Frame-theoretic analysis of oversampled filter banks}, IEEE Trans.
  Signal Process. \textbf{46} (1998), no.~12, 3256--3268.

\bibitem{Cvetkovic:2003a}
Z.~Cvetkovi\'c and J.~D. Johnston, \emph{Nonuniform oversampled filter banks
  for audio signal processing}, IEEE Speech Audio Process. \textbf{11} (2003),
  no.~5, 393--399.

\bibitem{Cvetkovic:1998a}
Zoran Cvetkovi\'c and Martin Vetterli, \emph{Oversampled filter banks}, IEEE
  Trans. Signal Process. \textbf{46} (1998), no.~5, 1245--1255.

\bibitem{dernecxxl15}
Olivier Derrien, Thibaud Necciari, and Peter Balazs, \emph{A quasi-orthogonal,
  invertible, and perceptually relevant time-frequency transform for audio
  coding}, Proc. EUSIPCO (Nice, France), IEEE, August 31 -- September 4 2015,
  pp.~804--808.

\bibitem{Donoho:1995a}
D.L. Donoho, \emph{De-noising by soft-thresholding}, IEEE Trans. Inf. Theory
  \textbf{41} (1995), no.~3, 613--627.

\bibitem{DBLP:journals/ijwmip/DorflerM14}
Monika D{\"{o}}rfler and Ewa Matusiak, \emph{Nonstationary gabor frames -
  existence and construction}, {IJWMIP} \textbf{12} (2014), no.~3.

\bibitem{Fickus:2013a}
Matthew Fickus, Melody~L. Massar, and Dustin~G. Mixon, \emph{Finite frames and
  filter banks}, Finite Frames, Applied and Numerical Harmonic Analysis,
  Birkhäuser Boston, Cambridge, MA, USA, 2013, pp.~337--379.

\bibitem{Flandrin:1999a}
P.~Flandrin, \emph{Time-frequency/time-scale analysis}, Wavelet analysis and
  its application, vol.~10, Academic Press, San Diego, CA, USA, 1999.

\bibitem{Gao:2014a}
Bin Gao, W.~L. Woo, and L.~C. Khor, \emph{Cochleagram-based audio pattern
  separation using two-dimensional non-negative matrix factorization with
  automatic sparsity adaptation}, J. Acoust. Soc. Am. \textbf{135} (2014),
  no.~3, 1171--1185.

\bibitem{Garofolo:1993a}
J.S. Garofolo, L.F. Lamel, W.M. Fisher, J.G. Fiscus, D.S. Pallett, and N.L.
  Dahlgren, \emph{{TIMIT} acoustic-phonetic continuous speech corpus
  {LDC93S1}}, Philadelphia: Linguistic Data Consortium, 1993.

\bibitem{Glasberg:1990a}
B.~R. Glasberg and B.~C.~J. Moore, \emph{Derivation of auditory filter shapes
  from notched-noise data}, Hear. Res. \textbf{47} (1990), 103--138.

\bibitem{Grochenig:1993a}
K.~Gr{\"o}chenig, \emph{Acceleration of the frame algorithm}, IEEE Trans.
  Signal Process. \textbf{41} (1993), no.~12, 3331--3340.

\bibitem{Gunawan:2010a}
Teddy~Surya Gunawan, Eliathamby Ambikairajah, and Julien Epps, \emph{Perceptual
  speech enhancement exploiting temporal masking properties of human auditory
  system}, Speech Commun. \textbf{52} (2010), no.~5, 381 -- 393.

\bibitem{Hansen:1998a}
John~HL Hansen and Bryan~L Pellom, \emph{An effective quality evaluation
  protocol for speech enhancement algorithms.}, Proc. ICSLP (Sydney,
  Australia), vol.~7, November 1998, pp.~2819--2822.

\bibitem{Hohmann:2002a}
V.~Hohmann, \emph{Frequency analysis and synthesis using a gammatone
  filterbank}, Acta Acust. united Ac. \textbf{88} (2002), no.~3, 433--442.

\bibitem{Holighaus:2013a}
N.~Holighaus, M.~D\"{o}rfler, G.~Velasco, and T.~Grill, \emph{A framework for
  invertible, real-time {constant-Q} transforms}, IEEE Audio, Speech, Language
  Process. \textbf{21} (2013), no.~4, 775--785.

\bibitem{Irino:2006b}
T.~Irino and R.~D. Patterson, \emph{A dynamic compressive gammachirp auditory
  filterbank}, IEEE Audio, Speech, Language Process. \textbf{14} (2006), no.~6,
  2222--2232.

\bibitem{Kovacevic:1993a}
Jelena Kova\v{c}evi\'{c} and M.~Vetterli, \emph{Perfect reconstruction filter
  banks with rational sampling factors}, IEEE Trans. Signal Process.
  \textbf{41} (1993), no.~6, 2047--2066.

\bibitem{6334422}
J.~Le~Roux and E.~Vincent, \emph{Consistent wiener filtering for audio source
  separation}, Signal Processing Letters, IEEE \textbf{20} (2013), no.~3,
  217--220.

\bibitem{Lin:2001b}
L.~Lin, W.H. Holmes, and E.~Ambikairajah, \emph{Auditory filter bank
  inversion}, Proc. ISCAS (Sydney, Australia), vol.~2, IEEE, May, 6--9 2001,
  pp.~537--540.

\bibitem{Lopez-Poveda:2001a}
E.~A. Lopez-Poveda and R.~Meddis, \emph{A human nonlinear filterbank}, J.
  Acoust. Soc. Am. \textbf{110} (2001), no.~6, 3107--3118.

\bibitem{Lyon:2010a}
R.F. Lyon, A.G. Katsiamis, and E.M. Drakakis, \emph{History and future of
  auditory filter models}, Proc. ISCAS (Paris, France), IEEE, June 2010,
  pp.~3809--3812.

\bibitem{majxxl1}
P.~Majdak, P.~Balazs, and B.~Laback, \emph{Multiple exponential sweep method
  for fast measurement of head related transferfunctions}, J. Audio Eng. Soc.
  \textbf{55} (2007), no.~7/8, 623--637.

\bibitem{majxxl10}
Piotr Majdak, Peter Balazs, Wolfgang Kreuzer, and Monika D\"orfler, \emph{A
  time-frequency method for increasing the signal-to-noise ratio in system
  identification with exponential sweeps}, Proc. ICASSP, 2011.

\bibitem{maba09}
D.~Marelli and P.~Balazs, \emph{On pole-zero model estimation methods
  minimizing a logarithmic criterion for speech analysis}, IEEE Trans. Audio,
  Speech, Language Process. \textbf{18} (2010), no.~2, 237--248.

\bibitem{Moore:2012a}
B.~C.~J. Moore, \emph{An introduction to the psychology of hearing}, sixth ed.,
  Emerald Group Publishing, Bingley, UK, 2012.

\bibitem{Necciari:2010a}
Thibaud Necciari, \emph{Auditory time-frequency masking: Psychoacoustical
  measures and application to the analysis-synthesis of sound signals}, Degree
  of {D}octor of {A}coustics, Aix-Marseille University, France, 2010.

\bibitem{Necciari:2013a}
Thibaud Necciari, Peter Balazs, Nicki Holighaus, and Peter S{\o}ndergaard,
  \emph{The {ERBlet} transform: An auditory-based time-frequency representation
  with perfect reconstruction}, Proc. ICASSP (Vancouver, Canada), IEEE, May
  2013, pp.~498--502.

\bibitem{ODonovan:2005a}
J.~J. O'Donovan and D.~J. Furlong, \emph{Perceptually motivated time-frequency
  analysis}, J. Acoust. Soc. Am. \textbf{117} (2005), no.~1, 250--262.

\bibitem{Oshaughnessy:1987a}
Douglas O'shaughnessy, \emph{Speech communication: human and machine},
  Addison-Wesley, 1987.

\bibitem{papkow13}
H{é}l{è}ne Papadopoulos and Matthieu Kowalski, \emph{Sparse and structured
  decomposition of audio signals on hybrid dictionaries using musical priors},
  J. Acoust. Soc. Am. \textbf{134} (2013), no.~1, 666--685.

\bibitem{Patterson:1992a}
Roy~D. Patterson, K.~Robinson, J.~Holdsworth, D.~McKeown, C.~Zhang, and Mike~H.
  Allerhand, \emph{Complex sounds and auditory images}, Auditory physiology and
  perception, Proceedings of the 9th International Symposium on Hearing
  (Oxford, UK), Pergamond, 1992, pp.~429--446.

\bibitem{ltfatnote022}
Zden\v{e}k Pr\r{u}\v{s}a, Peter~L. S{\o}ndergaard, Peter Balazs, and Nicki
  Holighaus, \emph{{LTFAT: A Matlab/Octave toolbox for sound processing}},
  Proc. CMMR (Marseille, France), October 2013, pp.~299--314.

\bibitem{ltfatnote030}
Zden\v{e}k Pr\r{u}\v{s}a, Peter~L. S{\o}ndergaard, Nicki Holighaus, Christoph
  Wiesmeyr, and Peter Balazs, \emph{{The Large Time-Frequency Analysis Toolbox
  2.0}}, Sound, Music, and Motion (Mitsuko Aramaki, Olivier Derrien, Richard
  Kronland-Martinet, and S{\o}lvi Ystad, eds.), Lecture Notes in Computer
  Science, Springer International Publishing, 2014, pp.~419--442.

\bibitem{Sadjadi:2015a}
Seyed~Omid Sadjadi and John~H.L. Hansen, \emph{Mean hilbert envelope
  coefficients (mhec) for robust speaker and language identification}, Speech
  Commun. \textbf{72} (2015), 138--148.

\bibitem{Schoerkhuber:2014a}
Christian Sch\"{o}rkhuber, Anssi Klapuri, Nicki Holighaus, and Monika
  D\"{o}rfler, \emph{A {Matlab} toolbox for efficient perfect reconstruction
  time-frequency transforms with log-frequency resolution}, Audio Engineering
  Society Conference: 53rd International Conference: Semantic Audio, AES,
  January 2014.

\bibitem{Shriberg:2007a}
Elizabeth Shriberg, \emph{Higher-level features in speaker recognition},
  Speaker Classification I (Christian Müller, ed.), Lecture Notes in Computer
  Science, vol. 4343, Springer Berlin Heidelberg, 2007, pp.~241--259.

\bibitem{sirdey:hal-00881707}
Adrien Sirdey, Olivier Derrien, and Richard Kronland-Martinet, \emph{{Adjusting
  the Spectral Envelope Evolution of Transposed Sounds with Gabor Mask
  Prototypes}}, Proc. DAFx-10 (Graz, Austria), September 2010, pp.~1--7.

\bibitem{balsto09new}
Diana~T. Stoeva and Peter Balazs, \emph{Invertibility of multipliers}, Appl.
  Comput. Harmon. Anal. \textbf{33} (2012), no.~2, 292--299.

\bibitem{Strahl:2009a}
Stefan Strahl and Alfred Mertins, \emph{Analysis and design of gammatone signal
  models}, J. Acoust. Soc. Am. \textbf{126} (2009), no.~5, 2379--2389.

\bibitem{Trefethen:1997a}
L.~N. Trefethen and D.~Bau~{III}, \emph{Numerical linear algebra}, SIAM,
  Philadelphia, PA, USA, 1997.

\bibitem{Unoki:2006a}
Masashi Unoki, Toshio Irino, Brian Glasberg, Brian C.~J. Moore, and Roy~D.
  Patterson, \emph{Comparison of the roex and gammachirp filters as
  representations of the auditory filter}, J. Acoust. Soc. Am. \textbf{120}
  (2006), no.~3, 1474--1492.

\bibitem{Vaidyanathan:1993a}
P.P. Vaidyanathan, \emph{Multirate systems and filter banks}, Electrical
  engineering. Electronic and digital design, Prentice Hall, Englewood Cliffs,
  NJ, USA, 1993.

\bibitem{Valero:2012a}
X.~Valero and F.~Alias, \emph{Gammatone cepstral coefficients: Biologically
  inspired features for non-speech audio classification}, IEEE Trans.
  Multimedia \textbf{14} (2012), no.~6, 1684--1689.

\bibitem{Venkitaraman:2014a}
Arun Venkitaraman, Aniruddha Adiga, and Chandra~Sekhar Seelamantula,
  \emph{Auditory-motivated gammatone wavelet transform}, Signal Process.
  \textbf{94} (2014), 608--619.

\bibitem{Vincent:2006a}
E.~Vincent, R.~Gribonval, and C.~Fevotte, \emph{Performance measurement in
  blind audio source separation}, IEEE Trans. Audio, Speech, Language Process.
  \textbf{14} (2006), no.~4, 1462--1469.

\bibitem{Zhao:2012a}
Xiaojia Zhao, Yang Shao, and DeLiang Wang, \emph{Casa-based robust speaker
  identification}, IEEE Trans. Audio, Speech, Language Process. \textbf{20}
  (2012), no.~5, 1608--1616.

\bibitem{Zilany:2006a}
Muhammad S.~A. Zilany and Ian~C. Bruce, \emph{Modeling auditory-nerve responses
  for high sound pressure levels in the normal and impaired auditory
  periphery}, J. Acoust. Soc. Am. \textbf{120} (2006), no.~3, 1446--1466.

\bibitem{Zwicker:1980a}
Eberhard Zwicker and E.~Terhardt, \emph{Analytical expressions for
  critical-band rate and critical bandwidth as a function of frequency}, J.
  Acoust. Soc. Am. \textbf{68} (1980), no.~5, 1523--1525.

\end{thebibliography}
\end{document}